\documentclass[prd,amsmath,aps,floats,amssymb, floatfix, superscriptaddress,nofootinbib,twocolumn,preprintnumbers]{revtex4-1}

\usepackage{amsmath,amssymb,natbib,latexsym,times,subfigure}

\usepackage{lineno}
\usepackage{makecell}

\usepackage[T1]{fontenc}
\usepackage{aecompl}

\usepackage{verbatim}
\usepackage{cprotect}
\usepackage{hhline,cuted}
\usepackage{ulem}

\usepackage{verbatim}
\usepackage[usenames,dvipsnames]{xcolor}
\usepackage{tabulary}
\usepackage{tabularx}
\usepackage{todonotes}
\usepackage{hyperref}

\newcommand\hmpcinv{\,h\,{\rm Mpc^{-1}}}

\allowdisplaybreaks
\begin{document}

\title{Constraining the Baryonic Feedback with Cosmic Shear Using the DES Year-3 Small-Scale Measurements}


\author{A.~Chen}
\affiliation{Department of Physics, University of Michigan, Ann Arbor, MI 48109, USA}
\affiliation{Kavli Institute for the Physics and Mathematics of the Universe (WPI), UTIAS, The University of Tokyo, Kashiwa, Chiba 277-8583, Japan}

\author{G.~Aric\`{o}}
\affiliation{Institute for Computational Science, University of Zurich, Winterthurerstrasse 190, 8057 Zurich, Switzerland}
\affiliation{Donostia International Physics Center (DIPC), Paseo Manuel de Lardizabal, 4, 20018, Donostia-San Sebasti\'an, Guipuzkoa, Spain}
\affiliation{Departamento de F\'isica, Universidad de Zaragoza, Pedro Cerbuna 12, 50009 Zaragoza, Spain}

\author{D.~Huterer}
\affiliation{Department of Physics, University of Michigan, Ann Arbor, MI 48109, USA}

\author{R.~Angulo}
\affiliation{Donostia International Physics Center (DIPC), Paseo Manuel de Lardizabal, 4, 20018, Donostia-San Sebasti\'an, Guipuzkoa, Spain}
\affiliation{IKERBASQUE, Basque Foundation for Science, 48013, Bilbao, Spain}

\author{N.~Weaverdyck}
\affiliation{Department of Physics, University of Michigan, Ann Arbor, MI 48109, USA}
\affiliation{Lawrence Berkeley National Laboratory, 1 Cyclotron Road, Berkeley, CA 94720, USA}

\author{O.~Friedrich}
\affiliation{Kavli Institute for Cosmology, University of Cambridge, Madingley Road, Cambridge CB3 0HA, UK}

\author{L.~F.~Secco}
\affiliation{Kavli Institute for Cosmological Physics, University of Chicago, Chicago, IL 60637, USA}

\author{C.~Hern\'{a}ndez-Monteagudo}
\affiliation{Department of Astrophysics Research, Instituto de Astrof\'{i}sica de Canarias, C\/V\'{i}a L\'{a}ctea, s\/n, E-38205, La Laguna, Tenerife, Spain}
\affiliation{Departamento de Astrof\'{i}sica, Universidad de La Laguna, Avenida Francisco S\'{a}nchez, s\/n, E-38205, La Laguna, Tenerife, Spain}

\author{A.~Alarcon}
\affiliation{Argonne National Laboratory, 9700 South Cass Avenue, Lemont, IL 60439, USA}

\author{O.~Alves}
\affiliation{Department of Physics, University of Michigan, Ann Arbor, MI 48109, USA}
\affiliation{Laborat\'orio Interinstitucional de e-Astronomia - LIneA, Rua Gal. Jos\'e Cristino 77, Rio de Janeiro, RJ - 20921-400, Brazil}

% ...

\collaboration{DES Collaboration}

\date{\today}

\label{firstpage}
\begin{abstract}
We use the small scales of the Dark Energy Survey (DES) Year-3 cosmic shear measurements, which are excluded from the DES Year-3 cosmological analysis, to constrain the baryonic feedback. To model the baryonic feedback, we adopt a baryonic correction model 
and use the numerical package \texttt{Baccoemu} to accelerate the evaluation of the baryonic nonlinear matter power spectrum. 
We design our analysis pipeline to focus on the constraints of the baryonic suppression effects, utilizing the implication given by a principal component analysis on the Fisher forecasts. Our constraint on the baryonic effects can then be used to better model and ameliorate the effects of baryons in producing cosmological constraints from the next generation large-scale structure surveys. We detect the baryonic suppression on the cosmic shear measurements  with a $\sim 2 \sigma$ significance. 
The characteristic halo mass for which half of the gas is ejected by baryonic feedback is constrained to be $M_c > 10^{13.2} h^{-1} M_{\odot}$ (95\% C.L.). The best-fit baryonic suppression is $\sim 5\%$ at $k=1.0 {\rm Mpc}\ h^{-1}$ and $\sim 15\%$ at $k=5.0 {\rm Mpc} \ h^{-1}$. Our findings are robust with respect to the assumptions about the cosmological parameters, specifics of the baryonic model, and intrinsic alignments.

\end{abstract}

\keywords{cosmology: cosmic shear; baryonic feedback}

\maketitle

\section{Introduction}

Baryons impact the density profiles of dark-matter-dominated cosmic structures on small spatial scales. As a consequence, they also affect the total-matter clustering signal. We call such baryonic physics impact on the total-matter clustering `baryonic feedback', incorporating many possible mechanisms like active galactic neuclei (AGN), gas cooling, metalicity evolution, etc. In most of the cases, AGN is the most important mechanism at the scale relevant to the large scale structure surveys, and it would lower the matter power by throwing a part of the baryonic matters out of the galaxy. While the effects of baryons are most noticeable in the clustering signal within individual halos, they extend out to the two-halo regime, on scales corresponding to a few megaparsecs. These effects thus complicate the cosmological inferences from surveys mapping out the clustering of cosmic structures. In order to mitigate this uncertainty in the cosmological analyses in the coming generation of large scale structure surveys, considerable effort has been undertaken to build reliable predictions for the baryonic feedback, including the adoption of the halo model \cite{mead2021hmcode}, principal component analysis on the baryonic suppression modes \cite{Huang:2018wpy}, and calibrated simulations \cite{SchneiderTeyssier2015, Arico2020}. In parallel, a growing number of analyses have been dedicated to assessing and validating these baryonic-modeling approaches \cite{maccrann2016inference, DES:2020rmk, Schneider2021, gatti2021cross, lee2022comparing, thiele2022percent, nicola2022breaking,Troster:2021gsz}. 

While the impact of baryons can be modeled with hydrodynamical N-body simulations \cite{Schaye2010, LeBrun2014,Schaye2015, McCarthy2017,Springel2018}, these results typically depend on the physics adopted in the simulations. Thus the inferred baryonic feedback  depends on the values of free parameters, which are in turn determined by a sub-grid recipe for baryonic physics. Because the hydrodynamical simulations are computationally very demanding, rerunning them for many baryonic scenarios quickly becomes prohibitive. Therefore, accurate yet efficient modeling of the effect of baryons on clustering remains a key challenge in cosmology. Addressing and solving this challenge will be required for upcoming surveys such as Euclid \cite{martinelli2020euclid}, the Rubin and Roman telescopes \cite{eifler2021cosmology}, the Dark Energy Spectroscopic Instrument (DESI) \cite{2020AAS...23544601F} and Subaru Prime Focus Spectrograph (PFS) \cite{takada2014extragalactic}.

`Baryonification' \cite{SchneiderTeyssier2015,Arico2020} is one such method that enables an efficient yet accurate modeling of the effects of baryons. 
This approach is based on the fundamental premise that the baryonic effects can be captured by shifting the position of particles in gravity-only N-body simulations. The shift is computed by means of parametrizing the difference between density profiles of cosmic structures with and without baryons. This introduces a few physically motivated free parameters which can be constrained with observations or hydrodynamical simulations. 

Our goal here is to apply the baryonification modeling to the data utilized in the Year-3 analysis of the Dark Energy Survey (DES) \cite{sheldon2020mitigating, sevilla2021dark,gatti2021dark, gatti2022dark}. These observations have a footprint of nearly 5,000 square degrees on the sky, and comprise the redshift and shape measurements of over 100 million galaxies, with the mean redshift $z=0.63$ \cite{secco2021dark}. In principle all of the key observations that comprise the `3x2pt'  DES Y3 analysis \cite{DES:2021wwk} --- galaxy clustering, galaxy-galaxy lensing, and cosmic shear --- would benefit from the baryonification analysis, as all three extend to scales potentially affected by baryons.
In this paper, however,  we only consider the DES Y3 observations of cosmic shear \cite{amon2021dark,secco2021dark}. We leave the application of baryonification to the full 3x2pt analysis to future work, because systematics other than the baryonic effect, for example the galaxy bias, are more predominant for the galaxy clustering and galaxy-galaxy lensing analysis.

A model dedicated to describe the baryonic effect on the large scale structure typically adds one or more free parameters to the cosmological parameter space, while enabling the extension of the clustering constraints to smaller scales. In this work among the first several adoptions of the baryonification model for a wide-field galaxy survey, we do a simpler analysis in order to study the effectiveness of this approach. We fix the cosmological parameters to the best-fit model derived in the standard analysis, then \textit{only} utilize the scales smaller than those used in the standard analysis in order to constrain the baryonification parameter(s). Therefore, instead of focusing on the cosmological parameters, we instead study the baryonic physics, measuring the amount of baryonic feedback in structure formation. The results obtained in this type of analysis can subsequently serve to provide a prior for the modeling of baryons in upcoming surveys, and thus help maximize the cosmological information from ongoing and future surveys such as DESI, Euclid, Rubin and Roman observatories, Hyper-Suprime Camera Survey (HSC), and Spherex.

The paper is organized as follows. In Sec.~\ref{sec: method} we describe the baryonification method and the corresponding numerical tools that we use. In Sec.~\ref{sec:analysis} we describe our analysis, and in Sec.~\ref{sec: results} we present its results. In Sec.~\ref{sec:disc} we discuss the results, and  compare them to others in the literature. We conclude in Sec.~\ref{sec:concl}. Additional information about our methods, results, and comparisons is available in the Appendices.

\section{Methodology}
\label{sec: method}
We model the matter power spectrum employing a series of Neural Network emulators from the BACCO Simulation project \citep{Angulo2020} (\texttt{Baccoemu}). Specifically, the matter power spectrum is decomposed in three different components: a linear part given by perturbation theory, a non-linear boost function, and a baryonic correction. The linear component is a direct emulation of the Boltzmann solver CLASS \citep{Lesgourgues2011}; it speeds up the calculations by several orders of magnitude while introducing a negligible error \citep{Arico2021}. The non-linear boost function is built by interpolating the output of  more than 800 simulations, obtained by scaling the cosmologies of six high-resolution $N$-body simulations of $\approx 2 {\rm Gpc}$ and $4320^3$ particles, using the methodology developed in \cite{AnguloWhite2010,Zennaro2019,Angulo2020,Contreras2020}. This algorithm manipulates a given simulation snapshot to mimic the expected particle distribution in a cosmological space that spans roughly the $10\sigma$ region around Planck 2018 best-fits \citep{Planck2018}. The parameter space includes dynamical dark energy and massive neutrinos, and models the power spectrum with an accuracy of $2-3\%$ \citep{Contreras2020,Angulo2020}. Finally, the baryonic correction is computed applying a baryonification algorithm \citep{SchneiderTeyssier2015, Arico2020} to the $N$-body simulations. The baryonification, or Baryon Correction Model (BCM), displaces the particles in a gravity-only simulation according to theoretically motivated analytical corrections \citep{SchneiderTeyssier2015, Arico2020} to model the impact of baryons on the density field. In the BCM framework, haloes are assumed to be made up of galaxies, gas in hydrostatic equilibrium, and dark matter. A given fraction of the gas is expelled from the haloes by accreting supermassive black holes, and the dark matter backreacts on the baryon gravitational potential with a quasi-adiabatic relaxation. The model employed has seven physically motivated free parameters, although we will show that varying just one parameter will be sufficient for our purposes. We refer the reader to Appendix \ref{app:BCM} for further details on the baryonification. By working at the field level, the BCM can provide predictions on multiple observable quantities, e.g. clusters' gas fraction from X-ray or kinetic Sunyaev-Zeldovich effect \citep{Giri2021,Schneider2021}. Moreover, the BCM has proven flexible enough to reproduce the 2-point and 3-point statistics of several hydrodynamical simulations \citep{Arico2020b}. The emulator that we employ fully captures the degeneracies between baryonic and cosmological parameters, while being accurate at several percent level \citep{Arico2020c}. 

Having emulated the nonlinear matter power spectrum with baryonic effects modeled by \texttt{Baccoemu}, we follow the same methodology as described in Section IV.B in Ref.~\cite{secco2021dark} to model the tomographic weak lensing two-point correlation functions. We are projecting the 3D matter power spectrum into 2D angular-space correlation functions, using the lensing kernel from the redshift-binned source galaxy samples. Hence we expect the baryonic suppression at small scales in the matter power spectrum to be reflected in the tomographic 2pt functions.


\section{Analysis Choices}\label{sec:analysis}


Our goal is to constrain the baryonic feedback using the DES Year-3 measurements of cosmic shear tomographic two-point correlation functions \cite{secco2021dark,amon2021dark} measured at small scales that were discarded in the standard cosmological analysis. We start with a Fisher forecast in Subsection \ref{sec:fisher} to inform how to reduce the dimensionality of the parameter space in the analysis. Then we specify the parameter priors and the nested sampling pipeline of our analysis in Subsection \ref{sec:priors} and \ref{sec:pipeline}. We discuss the possible systematics that could affect the baryonic feedback constraints in Subsection \ref{sec:systematics}, and at the end we finalize the blinding scheme in Subsection \ref{sec:blind} based on the considerations in this section. The real-data analysis pipeline is identical to the synthetic-data tests described in this section; the only difference is of course that fake data are replaced by real observations.


\subsection{Principal Component Analysis on Fisher Forecasts}
\label{sec:fisher}
The first choice to make in our analysis is to determine the baryonic parameter space that is sensitive to the precision of the measurements currently available to us. As recapped in Appendix \ref{app:BCM}, there are seven parameters introduced by the baryonic correction model adopted by \texttt{Baccoemu}. With the signal-to-noise of the small-scale cosmic shear measurements only, we are not likely to be able to constrain all of them.
Additionally, these unconstrained extra parameters can exacerbate convergence problems during the Monte Carlo sampling. 
We therefore need certain strategy to identify the subset of new parameters that are relevant to vary when analyzing an extended theoretical model, given the limited precision of data we have in hand. 

We introduce our innovative parameter space compression strategy as following. We define a metric $\mathcal{R}_{\rm FoM}\leq 1$ to quantify how well a multi-dimensional hypercube spanned by a subset of the parameters overlaps with the sub-parameter space best constrained by the data:
\begin{equation}
    \mathcal{R}_{\rm FoM}(\boldsymbol{\theta}) = \left. {\rm FoM_{\boldsymbol{\theta}}}\middle/ \prod_{i=1}^{N_{\boldsymbol{\theta}}} \sqrt{\lambda_i} \right.,
\end{equation}
where $\boldsymbol{\theta}$ is a subset of model parameters, $N_{\boldsymbol{\theta}}$ is the number of the parameters in this subset, $\lambda_i$ are the eigenvalues of the Fisher matrix in the decreasing order, and
\begin{equation}
    {\rm FoM_{\boldsymbol{\theta}}} = 1/\sqrt{\det{[(\mathcal{F}^{-1})|_{\boldsymbol{\theta}}]}}.
\end{equation}
Here we take the submatrix corresponding to $\boldsymbol{\theta}$ from the full Fisher matrix inverse to calculate the determinants.

We use two criteria to aid the identification of the parameters  $\boldsymbol{\theta}$ that are sensitive to the data:
\begin{itemize}
    \item \textbf{Criterion I}: $\mathcal{R}_{\rm FoM}(\boldsymbol{\theta}) \approx 1$. When $\mathcal{R}_{\rm FoM}(\boldsymbol{\theta}) $ approaches 1 from below, then the multi-dimensional hypercube spanned by $\boldsymbol{\theta}$ --- a subset of the model parameters --- overlaps with the space constrained by the first $N_{\boldsymbol{\theta}}$ principal components. 
    \item \textbf{Criterion II}: $\prod_{i=1}^{N_{\boldsymbol{\theta}}} \sqrt{\lambda_i}/\prod_{i=1}^{N} \sqrt{\lambda_i} \approx 1$, where $\lambda_i$ are normalized to $1$ for unconstrained (prior-dominated) principal components. This gives a measure of how much total information gain \textit{over the prior} is contained within just the first $N_{\boldsymbol{\theta}}$ principal components.
\end{itemize}
Both quantities featured in these two criteria are  $\leq 1$. When they approach unity simultaneously, then we can declare that the parameters not contained in $\boldsymbol{\theta}$ are \textit{in}sensitive to the data. We can thus justifiably vary $\boldsymbol{\theta}$ and fix all other parameters in the analysis.

Note that this whole argument is predicated on the assumption of a Gaussian posterior, which the Fisher matrix formalism assumes from the beginning. 

In our scenario of modeling the small scales of cosmic shear by introducing the BCM parameters, we first carry out the Fisher matrix calculation using the \texttt{fisher} routine of the \texttt{cosmosis} software. The Fisher matrix is defined as:
\begin{equation}
    \mathcal{F}_{ij} = \sum_{mn} \frac{\partial v_m}{\partial p_i} [C^{-1}]_{mn} \frac{\partial v_n}{\partial p_j} + [\mathcal{I}^{-1}]_{ij}.
\end{equation}
Here $v_m$ are the measured data points which are organized in a \textit{data vector}, $p_i$ are the model parameters, and $C$ is the measurement covariance matrix. Next, $\mathcal{I}_{ij}$ is the prior term, which is typically a diagonal matrix with elements $1/\sigma^2_i$ for uncorrelated priors, where $\sigma_i$ is the variance of the Gaussian prior of the $i$-th parameter. [For parameters on which we apply \textit{flat} priors, we calculate the equivalent Gaussian priors $\sigma_i$d, whose Gaussian variance scales with the flat prior range.]  When $\mathcal{F}_{ij}$ approaches $\mathcal{I}_{ij}$, the data are not providing information to the model parameters, and their constraints are dominated by the priors.

In the Fisher forecast, we vary the six cosmological parameters, 13 DES nuisance parameters, and seven baryonic parameters; see Table \ref{table:pars}. The six cosmological parameters that we vary are: matter and baryon densities relative to critical $\Omega_m$ and $\Omega_b$, amplitude of mass fluctuations $\sigma_8$, scaled Hubble constant $h$, and the (physical) neutrino density $\Omega_{\nu} h^2$. The detailed definitions of the 13 nuisance parameters (listed in Table \ref{table:pars} as intrinsic alignment, source photo-z shift, and shear calibration parameters) are given in Ref.~\cite{secco2021dark}, while the baryonic parameters are fully defined in Ref.~\cite{Arico2020b}. Note also that the characteristic masses, for example the halo mass scale that contains half of the total gas, $M_c$, are defined in units of $h^{-1} M_{\odot}$. This gives us a Fisher matrix with a total of 26 parameters.

We marginalize over all of the 13 DES nuisance parameters, as well as the three cosmological parameters $h, n_s$ and the sum of neutrino mass $M_{\nu}$, by dropping them from the inverse Fisher matrix. We do so because  cosmic shear measurements, which we are adopting here,  are known to be rather insensitive to all of these parameters. 

We then diagonalize the Fisher matrix in the remaining ten parameters to find the principal components in this final parameter space, 
which consists of the baryonic parameters and the cosmological parameters of interest, $\Omega_b, \Omega_m, \sigma_8$. Let us denote the eigenvalues in this 10D space, in descending order, as $\lambda_i$, $i = 1...10$, and the (normalized) principal components --- the eigenvectors --- as $p_i^{\theta_j}$. 

\begin{figure}[ht]
    \centering
    \includegraphics[width=0.48\textwidth]{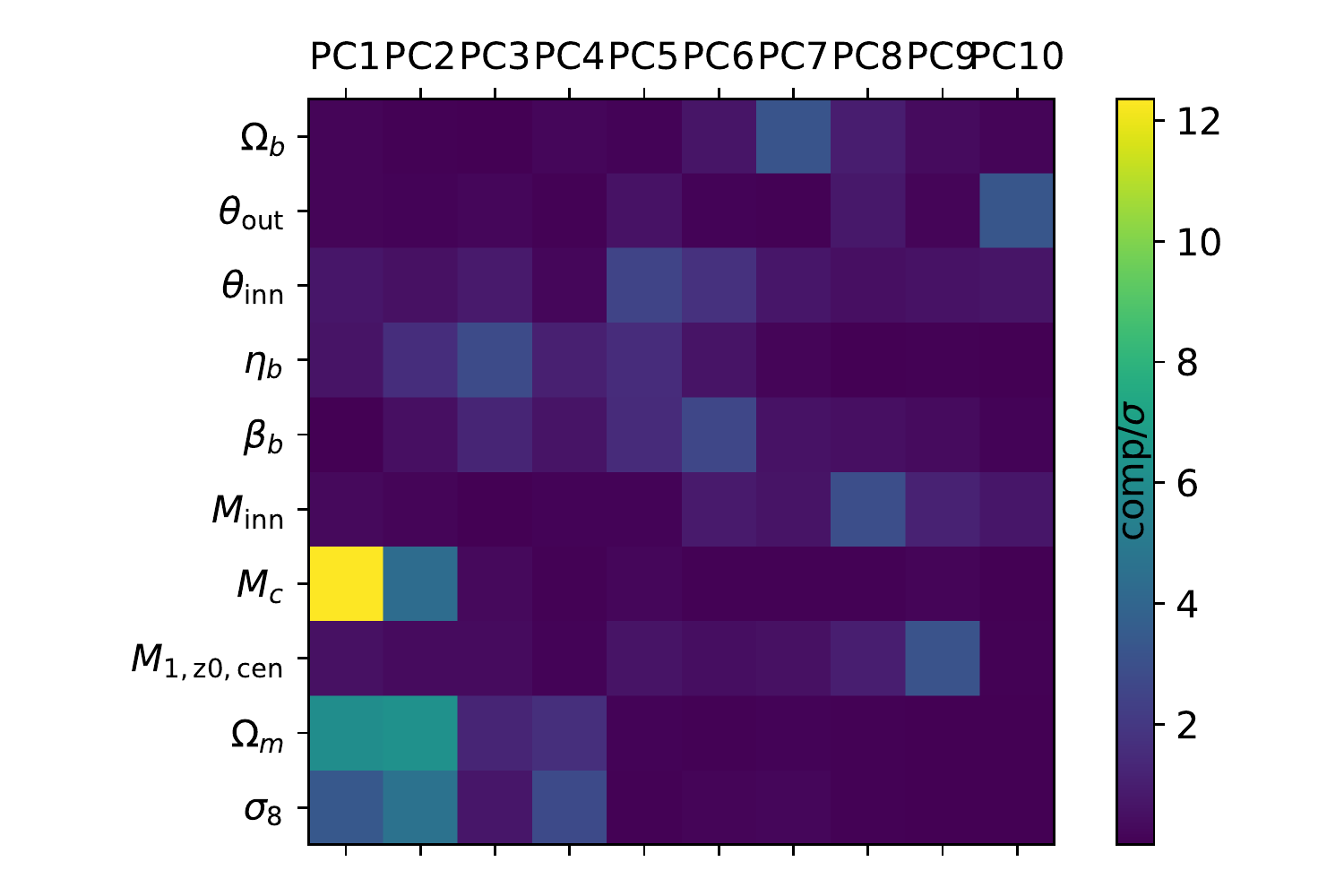}
    \caption{Color map of the value $|p_i^{\theta_j}*\sqrt{\lambda_i}|$, where $p_i^{\theta_j}$ is the coefficient of parameter $\theta_j$ in the principal component PC$_i$, and $\lambda_i$ is the eigenvalue of PC$_i$ for the ten-dimensional Fisher matrix. The Fisher matrix is calculated in the full parameter space with DES Year-3 small scales cosmic shear synthesized data vector, and is marginalized over the nuisance parameters and the unconstrained parameters $h, n_s$ and $M_{\nu}$; see text for details.}
    \label{fig:fisherpca}
\end{figure}

Figure \ref{fig:fisherpca} shows a color map in this 10D parameter space. The color is proportional to the quantities  $|p_i^{\theta_j}\times\sqrt{\lambda_i}|$, where $p_i^{\theta_j}$ is the coefficient of parameter $\theta_j$ in the principal component PC$_i$. These quantities combines the PCs' weights --- their eigenvalues --- with the coefficients of the parameters within that PC, to give an overall indication of how well the parameter is constrained by the data.  For example, $\log(M_c)$ has the largest coefficients of all parameters within the first (best-constrained) principal component, and can thus be reasonably expected to be the best-constrained single parameter in the full analysis. The first two principal components are dominated by $\log(M_c), \Omega_m$ and $\sigma_8$, indicating that these parameters are the best constrained by our data.

Next, we quantitatively check the two criteria we proposed above to investigate whether these three parameters are the only parameters constrained by the data instead of the priors. For the three-parameter subspace $(\log(M_c), \Omega_m, \sigma_8)$, Criterion I evaluates to
\begin{equation}
\begin{aligned}
    \mathcal{R}_{\rm FoM}(\log(M_c),\Omega_m, \sigma_8) &=\frac{{\rm FoM}_{\log(M_c),\Omega_m, \sigma_8}}{\prod_{i=1}^3 \sqrt{\lambda_i}}\\[0.2cm] 
    &= 0.836.
\end{aligned}
\end{equation}
In other words, when we choose to utilize three degrees of freedom to describe the constrained parameter space, the choice of the physically meaningful parameters $\log(M_c), \Omega_m$ and $\sigma_8$ can reproduce $83.6\%$ of the Figure of Merit of the more optimal but less interpretable choice of the first three principal components. 

For the Criterion II, when we normalize eigenvalues to $\lambda_i = 1$ for prior only principal components, we find
\begin{equation}
    \left. \prod_{i=1}^3 \sqrt{\lambda_i} \middle/ \prod_{i=1}^{10} \sqrt{\lambda_i} = 0.975 \right..
\end{equation}
This indicates that the PCs beyond the first three (so fourth, fifth, etc.\ PC) are almost fully prior dominated.  Hence, we conclude that the constrained parameter space for our small-scale cosmic shear analysis is almost completely spanned by the three parameters $\log(M_c), \Omega_m$ and $\sigma_8$, and we can fix the other cosmological and baryonic parameters. 

The parameter $M_c$ is defined in Ref.~\cite{Arico:2020yyf} as the halo mass scale 
that contains half of the total gas. In the same reference, they demonstrate that, among the seven BCM parameters, the baryon feedback suppression $S(k) = P_{\rm BCM}(k)/P_{\rm DMO}$ responds to the variation of $M_c$ most significantly; this agrees with our Fisher-forecast conclusions. Hence, in our real-data analysis, varying $\log(M_c)$ alone is analogous to measuring the amplitude of a specific pattern of baryon feedback, whose redshift and wavenumber dependence are motivated by theory and simulations. The priors on $\log(M_c)$ and other fixed BCM parameters will be presented in the next subsection. 

In conclusion, the Fisher PCA approach that we just described enabled us to determine the baryonic parameter space that can be constrained by the DES Y3 measurements. 

\subsection{Priors}
\label{sec:priors}
To get the best constraining power on the parameters that the small-scale cosmic shear analysis is sensitive to ( $\log(M_c), \Omega_m,$ and $\sigma_8$), we fix the other cosmological and baryonic parameters to the values based on best available information. We give the fixed cosmological parameters the mean values reported in the Planck-2018 TTTEEE+lowEE analysis \cite{Planck2018}. To the baryonic parameters other than $\log(M_c)$, we assign the values inferred from the power spectrum produced by the OWLS-AGN simulation at redshift $z=0$ \cite{Schaye2010}; see Table \ref{table:pars}. In the spirit of utilizing the available cosmological information to focus our constraining power on the baryonic parameters, we further apply the posterior in the $\Omega_m-\sigma_8$ space from the DES-Y3 3x2pt $\Lambda$CDM analysis as a part of our prior; we henceforth refer to this as the DES-Y3 prior. This prior, Gaussian but correlated in $\Omega_m$ and $\sigma_8$, captures information provided by large-scale analysis of weak lensing, galaxy clustering and galaxy-galaxy lensing. Because \texttt{Baccoemu} is trained around the best-fit of Planck cosmology, and there is a well-known $\sim\! 2 \sigma$ downward shift in the late universe $\sigma_8$ measurement compared to Planck, \texttt{Baccoemu} range covers only a half of our $\sigma_8$ prior at the higher value end, as illustrated in Figure \ref{fig:omsig8mc}. It is possible to cause some projection effect, which we leave to be taken care of in the future work, with an updated version of \texttt{Baccoemu} trained in larger spaces. 

\begin{table}[t]
\centering
\caption{Cosmological and nuisance parameters in our DES-Y3 small-scale cosmic shear analysis, and their priors. The `DES-Y3 3x2pt covariance.' label for $\Omega_m$ and $\sigma_8$ prior means that the posterior in this 2D parameters plane obtained from DES Y3 3x2pt analysis is applied as prior in our analysis (see the red contour in Figure \ref{fig:omsig8mc}). Our other cosmological parameters are fixed to the Planck bestfit values, and other BCM parameters are fixed to the bestfit to OWLS-AGN hydrodynamic simulations. The nuisance parameter priors (IA, photo-z shifts, and shear calibrations) are the same as DES Year-3 cosmic shear cosmological analysis.}
\begin{tabular}[t]{lccc}
\hline
Parameter & Prior\\
\hline
\multicolumn{2}{c}{\textbf{Cosmological}}\\
$\Omega_{\rm m}$& $\in [0.23,0.4]$, DES-Y3 3x2pt covariance. \\
$\sigma_8$ & $\in [0.73, 0.9]$, DES-Y3 3x2pt covariance. \\
$h$& 0.6727\\
$\Omega_{\rm b}$& 0.0493\\
$n_{\rm s}$& 0.9649\\
$\Omega_{\nu}h_0^2$& 0.00083 \\
\hline
\multicolumn{2}{c}{\textbf{Intrinsic Alignment}}\\
\multicolumn{2}{c}{TATT model}\\
$A_{1,IA}$ & flat (-5, 5)\\
$\alpha_{1,IA}$ & flat (-5, 5)\\
$A_{2,IA}$ & flat (-5, 5)\\
$\alpha_{2,IA}$ & flat (-5, 5)\\
$b_{\rm ta}$ & flat(0, 2)\\
\hline
\multicolumn{2}{c}{\textbf{Source photo-z shift}}\\
$\Delta z_s^1$ & Gauss (0.0, 0.018)\\
$\Delta z_s^2$ & Gauss (0.0, 0.015)\\
$\Delta z_s^3$ & Gauss (0.0, 0.011)\\
$\Delta z_s^4$ & Gauss (0.0, 0.017)\\
\hline
\multicolumn{2}{c}{\textbf{Shear calibration}}\\
$m_1$ & Gauss (-0.0063, 0.0091)\\
$m_2$ & Gauss (-0.0198, 0.0078)\\
$m_3$ & Gauss (-0.0241, 0.0076)\\
$m_4$ & Gauss (-0.0369, 0.0076)\\
\hline
\multicolumn{2}{c}{\textbf{BCM parameters}}\\
$\theta_{\rm out}$ & 0.419\\
$\theta_{\rm inn}$ & -0.702\\
$\eta_{b}$ & -0.248\\
$\beta_{b}$ & 0.321\\
$\log{M_{\rm inn}}$ & 13.0\\
$\log{M_c}$ & flat (12.0, 15.0)\\
$\log(M_{z0,\rm cen})$ & 10.4\\
\hline
\end{tabular}
\label{table:pars}
\end{table}%

\subsection{Pipeline}
\label{sec:pipeline}
We use \texttt{Baccoemu}\footnote{https://bacco.dipc.org/} \cite{Arico:2020yyf} to emulate the linear and nonlinear matter power spectrum with baryonic effects, as described in Section \ref{sec: method}. The maximum wavenumber encoded by the emulator goes up to $k=5.0$, and beyond this scale we linearly extrapolate the logarithm of the matter power spectrum to high-k for 2D projection purpose. We use the data vector of cosmic shear measurements in configuration space, $\xi_{\pm}$, only at small scales. Namely, we use the same scale cuts as the fiducial DES Year-3 cosmic shear analysis, but in the opposite way, adopting only the data points at angles smaller than the scale cuts. Because DES cosmic shear scale cuts are determined by minimizing the effects of baryonic feedback \cite{secco2021dark,amon2021dark}, adopting the complementary scale cuts lets us utilize the data that are the most sensitive to the baryonic feedback. With this removal of the large scales used in the cosmology analysis adopted, we have 173 data points (measurements of $\xi_{\pm}$). As shown in Figure \ref{fig:xipxim}, on these small scales there are many more $\xi_-$ data points than $\xi_+$, which is exactly the opposite from the situation in the standard cosmological analysis. This is because the structure of the $\xi_-$ kernel makes it more significantly based on small scales, and hence affected by the baryonic effects. Measurements of $\xi_-$ thus provide particularly valuable information on the BCM parameters.

We use \texttt{Cosmosis} \cite{cosmosis} \footnote{https://github.com/joezuntz/cosmosis}, \texttt{Polychord} \cite{handley2015polychord}, \texttt{Camb} \cite{lewis2011camb}, \texttt{GetDist} \cite{lewis2019getdist} for the nested sampling and the analysis pipeline.

\subsection{Systematics}
\label{sec:systematics}
Baryonic feedback is an important effect at relatively small,  nonlinear spatial scales, but it is by no means the \textit{only} effect at small scales. Hence when using the small scale cosmic shear measurements to constrain the baryonic effects, we need to ensure that the systematic uncertainties introduced by other small-scale effects are under control. Here we investigate the systematics related to the intrinsic alignment and nonlinear clustering. We also discuss the systematics induced by possible incorrect assumptions on the cosmological parameters, then conclude with a strategy to balance the constraining power and the bias on the baryonic parameter $\log(M_c)$. 

For the investigation of several systematics that are fairly subdominant and not marginalized by modeling, we use the following strategy: we generate synthetic data vectors contaminated by certain systematics, then carry out the standard analysis by simply ignoring these systematics. We compare the posterior of $M_c$, the parameter that we concern the most in our analysis, between the baseline analysis and the contaminated data vector analysis. We claim that the systematics is under control when the shift in $M_c$ is $<0.2 \sigma$.

\begin{figure}[ht]
    \centering
    \includegraphics[width=0.48\textwidth]{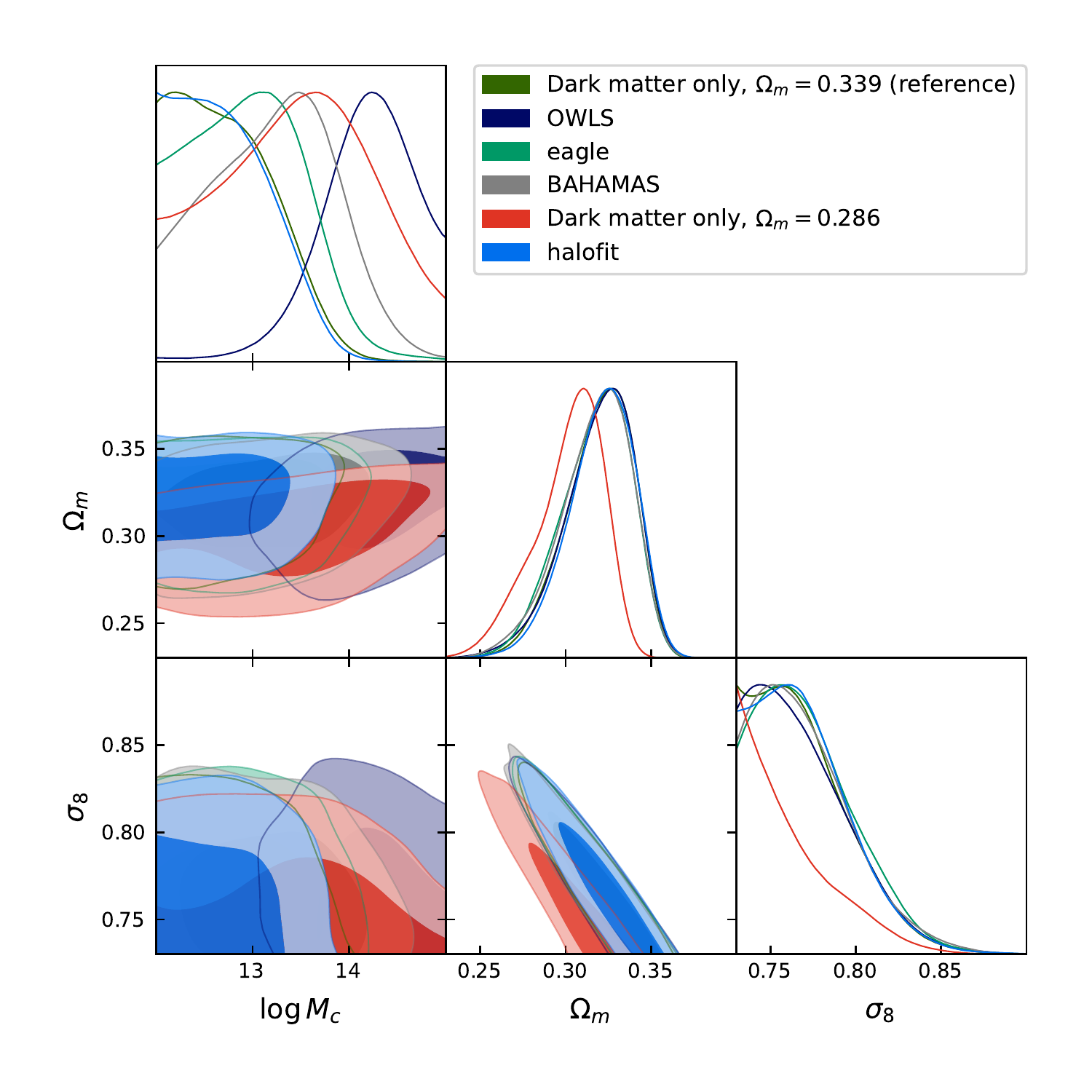}
    \caption{ Test results using the baseline baryonification analysis pipeline on the synthetic data vectors. We adopt the dark-matter-only matter power spectrum (corresponding to no baryonic suppression) as our {\it a priori} fiducial reference. We then generate the matter power spectra suppressed by the redshift-dependent baryonic effect measured in OWLS-AGN \cite{Schaye2010}, BAHAMAS\_T7.6\_WMAP9 \cite{McCarthy2017}, and eagle \cite{Schaye2015} simulations \cite{Huang:2018wpy,vanDaalen:2019pst} (corresponding to decreasing amplitude of baryonic suppression). The figure shows that our analysis pipeline successfully captures the relative amplitudes of the baryonic effect between different simulations in $\log{M_c}$ posteriors. We also test the synthetic dark-matter-only data vector with a lower $\Omega_m$ than our parameter prior, and this leads to  $\log{M_c}$ that is biased high, as shown in the contour labeled with $\Omega_m=0.286$. Lastly, we replace the data-vector-generating nonlinear module, switching from \texttt{Baccoemu} to \texttt{halofit}, and find that their difference is not introducing statistically significant bias in the parameter space that we are interested in.}
    \label{fig:synthetic}
\end{figure}

\subsubsection{Intrinsic Alignments}
The ellipticity of the observed galaxies is induced by either the weak lensing of the background galaxies, or else by the intrinsic alignments (IA) caused by the tidal gravitational force from cosmic structures 
Intrinsic-alignment auto and cross correlations with shear are expected to have a larger effect at smaller scales. Hence we adopt a beyond-linear, perturbative-theory model to predict the intrinsic alignment in our analysis  --- the Tidal Alignment and Tidal Torquing (TATT) model \cite{Blazek:2017wbz}. The precise range of scales over which the TATT model is accurate is still under investigation, but here we argue that a straightforward application of TATT is sufficient for us for two reasons. First, TATT is quite flexible, as it introduces up to five nuisance parameters to capture the IA power. Second, at very small scales ($\lesssim 2$Mpc) where TATT may start to become less accurate, the statistical errors of the DES cosmic shear measurements start to rapidly increase. Therefore, even  though some nonlinear IA features may not be captured by the parameter space of TATT model, they are unlikely to affect our results significantly. 

With the reasoning above, we carry out our real-data analysis marginalizing over the TATT model parameters for the intrinsic alignment. After unblinding, we investigate the possible degeneracies between the IA parameters and the baryonic suppression, as discussed in Appendix \ref{app:IA}. We confirmed, based on the contours in Figure \ref{fig:tatt}, that: 1.\ The TATT model parameters are not correlated with $M_c$; 2.\ In our BCM analysis the constraint on the TATT parameters is consistent with DES Year-3 3x2pt and cosmic shear 1x2pt cosmological analysis results. We thus conclude that the intrinsic alignment is not biasing our baryonic physics constraints. The caveat of the above argument is that we trust the degrees of freedom introduced by TATT model to be able capture the IA features to the accuracy required by the quality of our small-scale data.

\subsubsection{Nonlinear Matter Power Spectrum}
In our fiducial analysis pipeline,  nonlinear physics is modeled by \texttt{Baccoemu}. However, there still remain different choices that one can make in  modeling the nonlinear clustering of dark matter alone; see e.g.~\cite{martinelli2020euclid}. To address this,  we ran our baseline analysis on the synthetic data vector generated by an alternative nonlinear matter power spectrum model. For this alternative,  we chose \texttt{takahashi-halofit} \cite{takahashi}. As shown in Figure \ref{fig:synthetic}, the posteriors on $\log(M_c), \Omega_m$ and $\sigma_8$ are almost indistinguishable from the baseline case, with the tension between two posteriors being $<0.02\sigma$, so we conclude that the nonlinear-modeling uncertainty will not be an issue in our analysis.

\subsubsection{Cosmological Model Assumptions}

As discussed in Sec.~\ref{sec:priors},  we fix many of the cosmological parameters, an set priors on  additional few, in order to focus on the constraints  on the baryonic feedback. A natural concern in such an approach is the possible bias in our results introduced by incorrect assumptions on the cosmological model (relative to the ground truth, whatever it may be). To address these concerns, we perform a validation test with an alternative value of a key cosmological parameter. Specifically, we run a chain on dark-matter-only synthetic data vector centered at the value of the matter density that is at the \textit{lower end} of the 95\% credible-level constraint in the DES Year-3 3x2pt analysis. That is, given the 95\% C.L. DES Year-3 constraint  $\Omega_m\in [0.286, 0.390]$, we adopt $\Omega_m=0.286$, thus replacing our baseline which is the DES-Y3 central value, $\Omega_m = 0.339$. As illustrated by the red contour in Figure \ref{fig:synthetic}, lower $\Omega_m$ value shifts the marginalized $\log(M_c)$ posterior away from its baseline of $\log(M_c)=12.0$ to a higher value in the $\log(M_c)=13.5$-$14.0$ range, with $\sim 0.8 \sigma$ significance. Fortunately, such a scenario leaves an unambiguous additional signature, which is a shift, relative to the prior, in the $\Omega_m-\sigma_8$ constraint; see the red contour relative to the others in this plane in Figure \ref{fig:synthetic}. Therefore, one thing to monitor will be the comparison of the \textit{small-scale} $\Omega_m-\sigma_8$ posterior and that obtained in the standard cosmological analysis that utilizes large scales. Any mismatch between those two may indicate a  possible bias in the inferred baryonic parameter $\log(M_c)$ as well. We will see below that our analysis analysis does not show indications any such shift. 

\subsubsection{Higher-order Cosmic Shear}
\begin{figure}[ht]
    \centering
    \includegraphics[width=0.48\textwidth]{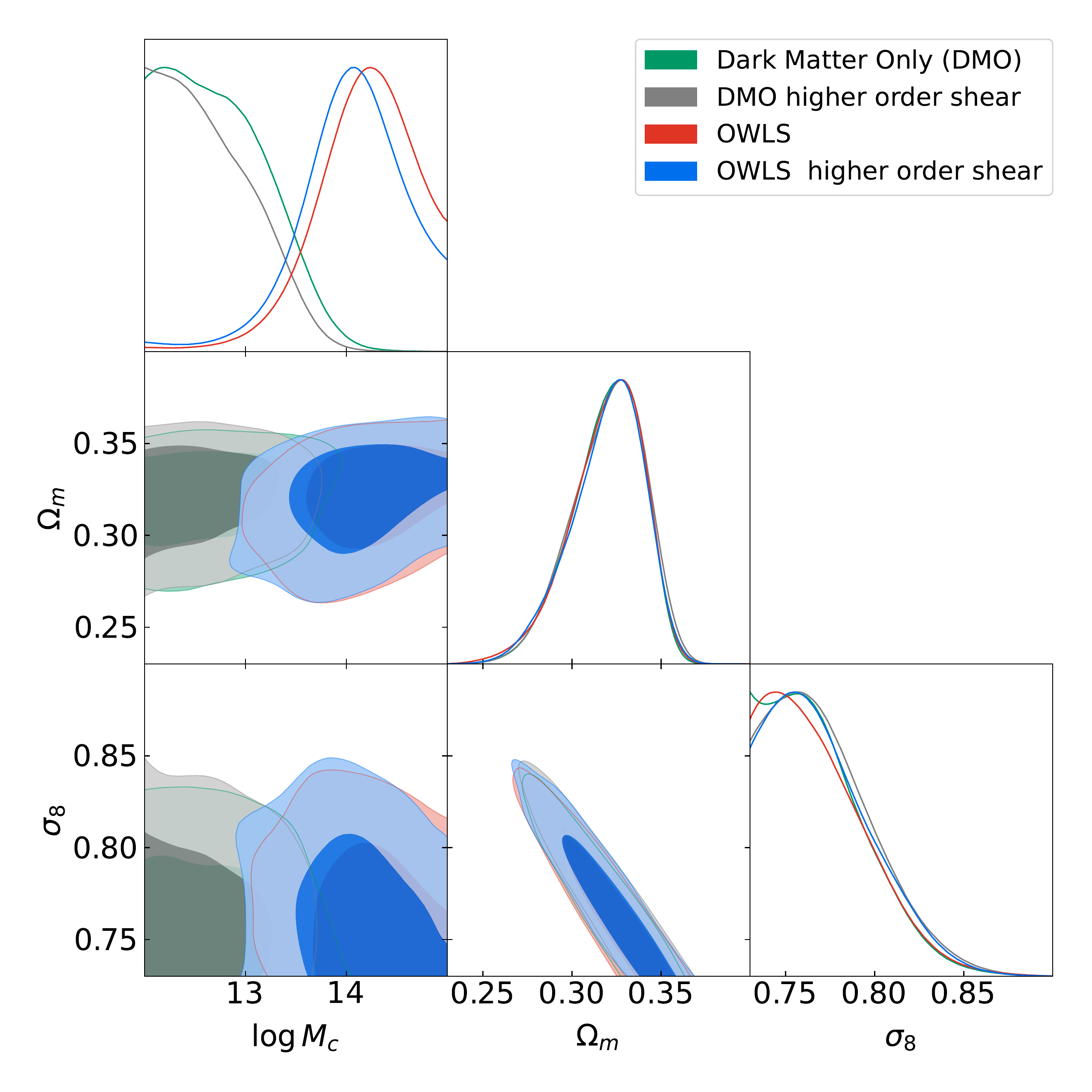}
    \caption{Synthetic data vector tests for higher order shear effects. The DMO and OWLS contours are the same in the Figure \ref{fig:synthetic}, and the two higher order shear synthetic data vectors are generated on the base of DMO and OWLS ones multiplying the ratio depicting the higher order shear effects from Figure 5 of \cite{secco2021dark}. The effect of higher order shear is indicated by the deviation from green contour to grey contour, and from red contour to blue contour. In neither cases higher order shear effects cause large shift in $\log{M_c}$, and the shift direction is always toward smaller value of $\log{M_c}$, i.e. less possibility of a fake detection of the baryonic suppression.}
    \label{fig:higherordershear}
\end{figure}

Higher-order cosmic shear corrections, including the reduced shear \cite{Dodelson:2005ir} and source magnification \cite{Schneider:2001af}, have been studied in the DES Year-3 methodology paper \cite{krause2021dark,secco2021dark}. As shown in the Figure 5 of \cite{secco2021dark}, systematics due to higher-order cosmic shear effects are generally subdominant to the baryonic suppression. Assuming that such effects are roughly cosmology-independent, we apply the higher-order shear effects depicted by the purple dotted line in Figure 5 of \cite{secco2021dark} on our dark-matter-only and OWLS synthetic data vector. The bias introduced by not including such effect in our modeling pipeline are $<0.05\sigma$ and $<0.2\sigma$ for DMO and OWLS cases as shown in Figure \ref{fig:higherordershear}.  Hence, we conclude that higher-order corrections to shear are not a concern.

\subsection{Blinding }
\label{sec:blind}
To avoid confirmation bias, we blind our results --- that is, we do not reveal our principal results until we have finalized our analysis and modeling criteria and choices. Our decision to blind is motivated by the increasing realization that complex cosmological analyses require at last some level of blinding in order to prevent unintended, subjective factors in biasing the analysis results \citep{muir2020blinding}. Note that every aspect of our real-data analysis that leads to the results presented in Sec.~\ref{sec: results} is the same as in our synthetic data tests, and that we did not alter any analysis choices after unblinding. At the same time, we must keep in mind that the DES Year-3 cosmology analysis using the large scales has already been done and is publicly available, and thus we are \textit{not} blind to the analysis choices that have been made there and that influenced our choices in this work.

Recall, our key results will be the posteriors and other statistical measures in $\log(M_c), \Omega_m$ and $\sigma_8$. It is the constraints on these parameters that we want to blind until our analysis choices have been finalized. We now summarize our blinding procedure. 

Before unblinding, we calculate the posterior predictive distribution (PPD) p-value of the BCM model; for details, see Ref.~\cite{DES:2020lei}. The goal of this step is to guarantee that our model represents a  reasonable description of the data. The PPD p-value characterizes the probability that the $\Delta \chi^2 = (\bold{D}-\bold{M})^{\rm T}C^{-1}(\bold{D}-\bold{M})$, evaluated between the data $\bold{D}$ and the theory prediction $\bold{M}$ for some values of the parameters, is smaller than the $\Delta \chi^2$ evaluated between a multi-variate Gaussian \textit{realization} of the data and the noiseless theory data vector. The latter quantity should obey the chi-squared distribution with the degrees of freedom equal to the number of data points, so we calculate PPD p-value as:
\begin{equation}
    p^{\rm PPD} = \sum_i (1-F_{[\rm \#\ of\ pts]}(\Delta\chi^2_i)) \times w_i
    \label{eq: ppd}
\end{equation}
where $F_{k}(x)$ is the cumulative distribution function of a chi-squared distribution with $k$ degrees of freedom, $\Delta\chi^2_i$ is evaluated between the real data and the theory prediction at $i$-th sample in the MCMC chain, and $w_i$ is the weight of the sample. The passing criterion for unblinding is $p^{\rm PDD}>0.01$. All of our real-data chains pass this criterion; the specific values of $p^{\rm PPD}$ are reported in the results section below.

\begin{figure*}
\includegraphics[width=0.49\textwidth]{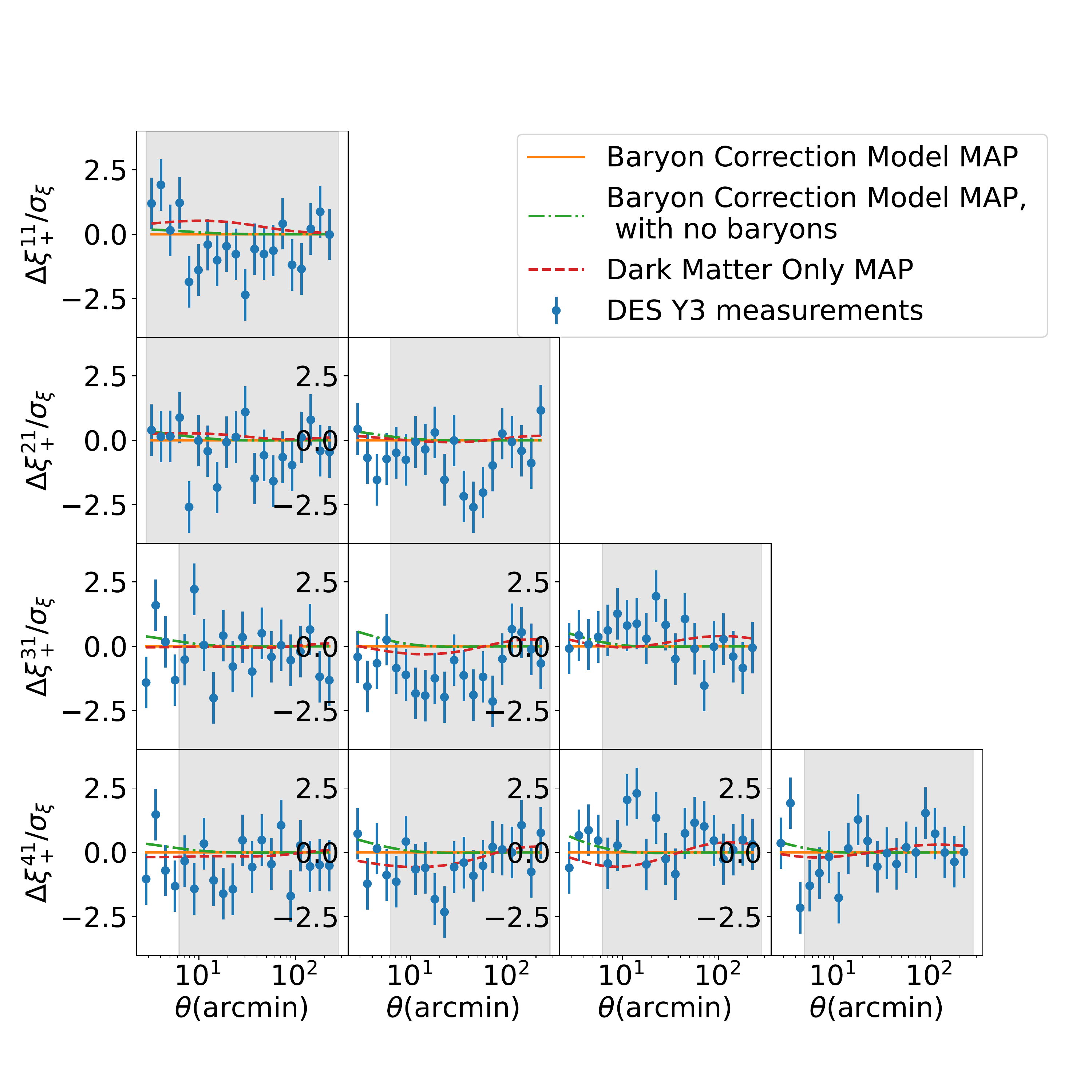}
\includegraphics[width=0.49\textwidth]{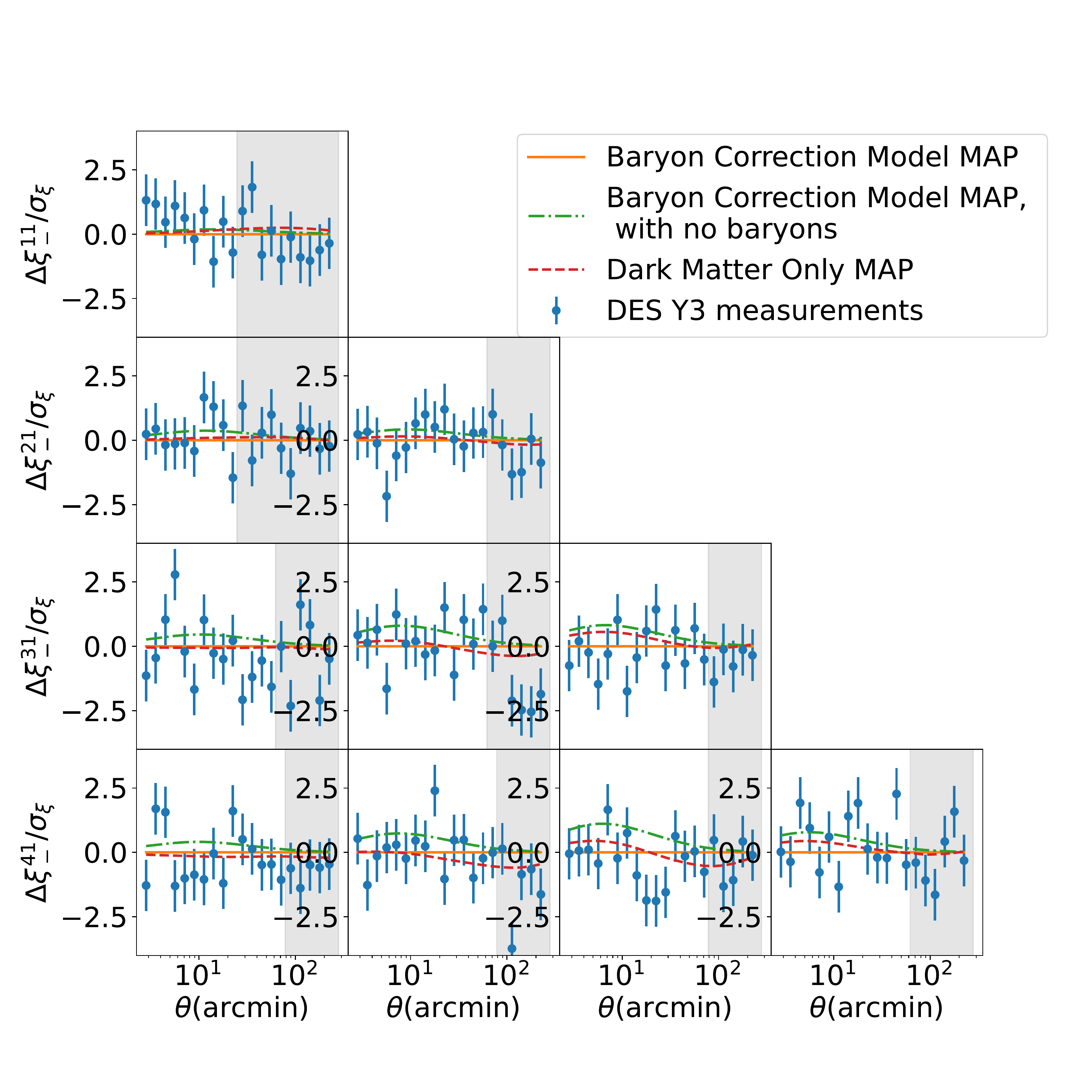}
\caption{
 Tomographic cosmic-shear two-point correlation functions, $\xi_+$ (left panel) and , $\xi_-$ (right panel) \cite{amon2021dark}. All of the curves --- both theory and data --- are shown relative to the bestfit prediction under BCM, and are divided by the observational errors.   The blue points are DES Year-3 measurements. Orange curves, equal to precisely zero, correspond to our best-fit BCM model's prediction. The green curves are the predictions keeping everything the same as the orange curves, but with the baryonic suppression artificially turned off. The red dashed curves are the dark-matter-only bestfit prediction. The grey regions show the scales \textit{not} used in our analysis. }
\label{fig:xipxim}
\end{figure*}

Having passed the PPD criterion, we also plot the maximum \textit{a posteriori} (MAP) theoretical data vector from the chains with the measured data points in Figure \ref{fig:xipxim} to further confirm that the MAP of the chains reasonably capture the measurements. The cosmic shear measurements denoted by the blue dots are well within the observational uncertainty around the MAP best-fit theory prediction of our baryonification model, denoted by the orange horizontal line.


\section{Results}
\label{sec: results}

As mentioned above, we pass the unblinding criteria that were pre-specified for our analysis. Specifically, we find a good consistency between the data and the baseline baryon correction model (and baseline analysis choices), with PPD p-value $p= 0.50$ (see Eq.~\ref{eq: ppd}). We thus unblind the analysis at this point.

\begin{figure}[ht]
    \centering
    \includegraphics[width=0.48\textwidth]{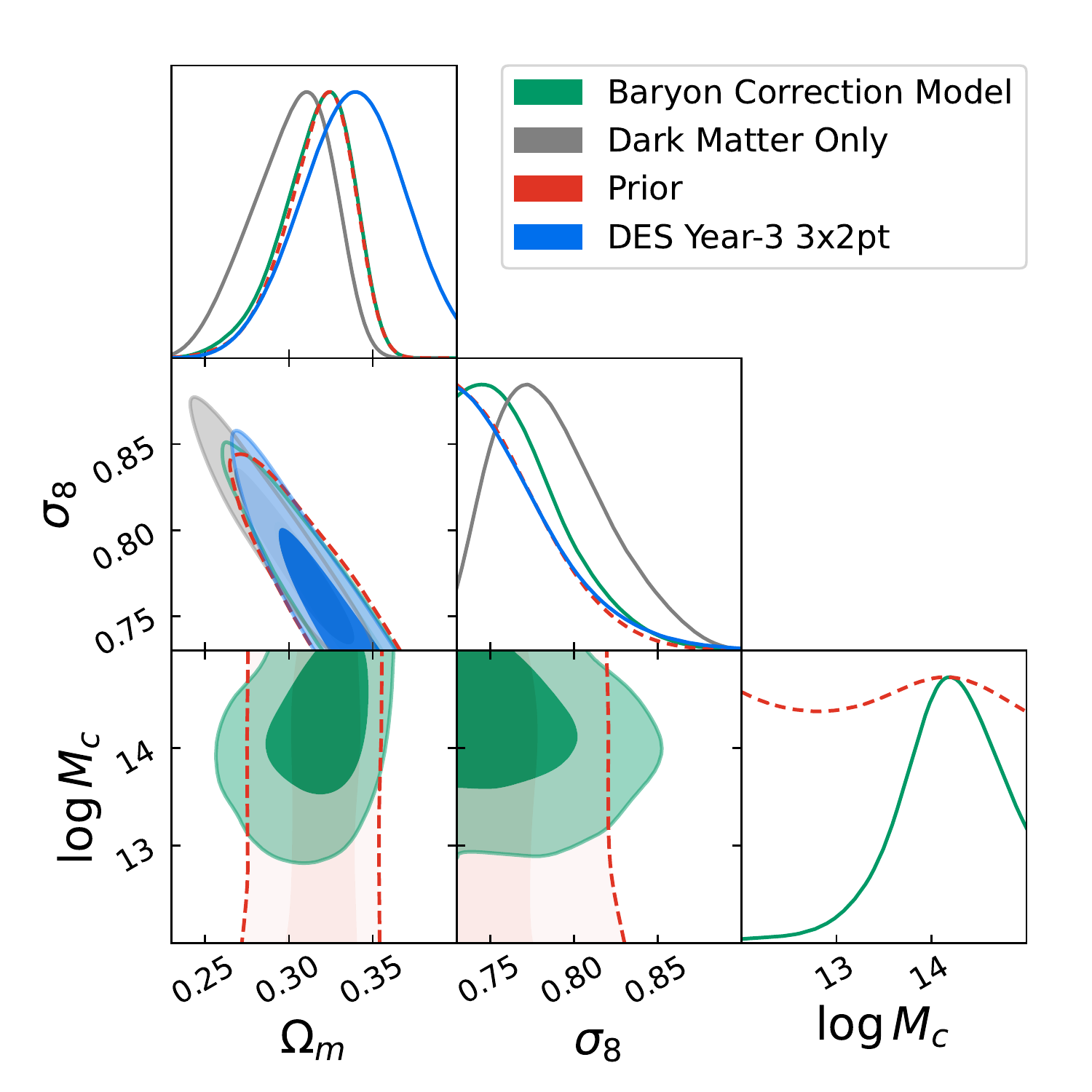}
    \caption{The constraints on $\Omega_m, \sigma_8$ and $\log{M_c}$. Note that the dark-matter-only small-scale cosmic shear analysis and DES Year-3 3x2pt analysis do not have $\log{M_c}$ in their models. The prior in the $\Omega_m$-$\sigma_8$ plane is taken to be the DES Year-3 3x2pt parameter posterior covariance. The shift in 1D marginalized $\Omega_m$ probability distribution in our BCM baseline away from the DES Year-3 3x2pt analysis is 
    caused by the lower limit in $\sigma_8$ due to the limited sampling range of the emulator.}
    \label{fig:omsig8mc}
\end{figure}

The main result is the constraint in the $(\Omega_m, \sigma_8, \log(M_c))$ space shown in Figure \ref{fig:omsig8mc}. We detect the $\log(M_c)$ value to be away from the lower bound of $\log(M_c)=12.0$ which obtains in the dark-matter-only limit. We find:
\begin{alignat}{2}
    \log(M_c) & =14.12^{+0.62}_{-0.37} & \quad \mbox{68\% C.L.,}\\[0.1cm]
      \log(M_c)  & > 13.2 & \quad \mbox{95\% C.L.} .
\end{alignat}
As an illustration of the suggested magnitude of the baryonic feedback on the cosmic shear two-point correlation functions, Figure \ref{fig:xipxim} shows comparison of the MAP result using theory with the BCM baseline analysis (orange; equal to precisely zero in the Figure) and theory without BCM (green), using the same parameters. The suppression of the theory with baryonic feedback --- so, where orange curves are lower than green curves --- is noticeable at the small scales of $\xi_-$, especially in the higher redshift bins. This trend can be explained by the combined effect of the increase in the $\xi_-$ amplitude toward higher redshift, wider coverage of the lensing kernel (longer light path), and shrinking of the measurement uncertainty. 

In Appendix \ref{app:effectiveredshift}, we show that the effective redshift of our baryonic effect constraint is relatively low, $z_{\rm eff} \approx 0.21$. This effective redshift is defined as the value at which our small-scale cosmic shear data vector responds most strongly to the redshift-localized BCM evaluated at that redshift. The low $z_{\rm eff}$ could be caused by the fact that the cosmic shear characterizes an integrated effect over the light path traveled from the source galaxy, so the effects that kick in at low redshifts are probed by multiple tomographic redshift bins. Another possibility is that the baryonic feedback is intrinsically strong at lower redshifts, but due to the integral nature of the lensing kernel, we cannot confirm this hypothesis from our analysis. 

\section{Discussion}\label{sec:disc}

\subsection{Model Comparison with a Dark Matter Only Universe}
The constraint on $\log(M_c)$ reported above disfavors the hypothesis of the dark-matter-only nonlinear matter power spectrum. Namely, $\log(M_c) > 13.2$ suggests the presence of the baryonic suppression mode at small scales of scales probed by cosmic shear. In this section, we evaluate the statistical significance of this finding by carrying out a more detailed comparison between the cosmological models with and without baryons. In the following text, the dark matter only (DMO) cosmology refers to a cosmology with no baryonic effect, hence all the masses are effectively dark matters which only interact through gravity. When DMO is used on a simulation, it refers to the gravity-only N-body simulations.

We calculated several popular information criteria as metrics for the model comparison in the Table \ref{table:ICs}. Their definitions are formulated in Table \ref{table:ICdefinitions} of Appendix \ref{app:ICs}. In general all the information criteria utilize the idea that the improvement in the fitting to the measurements, i.e. the decrease of $\chi^2$, should be punished by extra degrees of freedom of the model. Specifically, each information criterion takes a metric of the $\chi^2$ (minumum or average), and a definition of the number of degrees of freedom, and combines them into one quantity. We use two alternate ways to measure the number of degrees of freedom $k$ in a model: the Bayesian Model Dimensionality, BMD \cite{Handley:2019pqx}, and the simple counting of the free model parameters, $N$. The latter should provide the most conservative way of interpreting our findings, as the simple parameter count  corresponds to the maximum possible number of degrees of freedom of a model. Due to the presence of priors, the effective degrees of freedom of a model $k$ is always smaller than $N$. The difference in the counting of DMO and BCM model parameters, $\Delta N$, is one, corresponding to the parameter $M_c$. Despite the details above, in all the statistical tests listed in Table \ref{table:ICs}, baryonification is preferred, at very strong (${\rm XIC}<-3.5$, where XIC stands for a certain information criterion) or moderate ($-2.3<{\rm XIC}<-1.2$) level as evaluated on Jeffreys' scale \cite{jeffreys1961theory,robert2009harold,Nesseris:2012cq}.

We now provide estimates of the preference for the baryon correction model. Assuming that the exponential of information criteria reflects the ratio of the two hypotheses:
\begin{itemize}
\item H0: We live in dark-matter-only (DMO) universe
\item H1: We live in a universe with baryons, and we need an additional parameter $M_c$ to describe them,
\end{itemize} 
We convert the probability $p$ preferring H1 into the easy-to-gauge number of standard deviations (`sigmas'), $z$:

\begin{equation}
    z = \sqrt{2}\ {\rm erf}^{-1}(p)
\end{equation}
The bottom row of the Table \ref{table:ICs} shows the converted number of sigmas. It shows that, in all cases, the hypothesis H1 with baryons is preferred at evidence that ranges from $1.4\sigma$ to $2.7\sigma$.

Note that there are differences between the information criteria calculated using the Bayesian Model Dimensionality (BMD) \cite{Handley:2019pqx} and using the parameter counting $N$. The strong preference for the model with baryons using the BMD largely comes from this decrease of BMD in the baryon model relative to the dark-matter only case (note the negative value in the fourth row, fourth column of Table \ref{table:ICs}). This decrease of the baryon models' degrees of freedom is counter-intuitive, because we actually \textit{add} one degree of freedom when we go from DMO to the baryon model. The reported decrease of model dimensionality for the baryon case is likely telling us that the data fit the baryon model's features better on average. Note that BMD roughly corresponds to the variance of $\chi^2$ for the sampled points in the chain (see the formula for BMD in Table \ref{table:ICdefinitions}). The reported decrease in BMD therefore suggests that there exists a  locus in the parameter space in which the data vector prefers to settle. 

{\renewcommand{\arraystretch}{1.3}
\begin{table*}[ht]
\centering
\caption{Model comparison metrics between the baryonification and dark matter only model. We calculate Akaike Information Criterion, Bayesian Information Criterion, and Deviance Information Criterion using both the Bayesian Model Dimension (BMD) \cite{Handley:2019pqx} and the naive parameter counting (N). The bottom row converts the difference in the information criteria between the BMD and DMO model into a significance for the presence of the baryonic parameter $M_c$ quoted in `sigmas'; see text for details.}
\begin{tabular}[t]{lcccccccccc}
\hline
 & $\chi^2_{\rm min}$ & $\langle \chi^2 \rangle$ & BMD & N & AIC(k=BMD) & AIC(k=N) &BIC(k=BMD) & BIC(k=N) & DIC(k=BMD) & DIC(k=N)\\
\hline
Baryonification (BCM) & 163.6 & 172.3 & 5.5 & 16 & 174.5 & 195.6 & 175.8 & 199.4 & 183.2 & 204.3\\
\hline
Dark Matter Only & 168.4 & 176.2 & 6.0 & 15 & 179.3  & 197.4 & 180.8 & 200.9 & 188.2 & 206.2\\
\hline
BCM - DMO & $-3.8$ & $-3.9$ & $-0.5$ & $1$ & $-4.8$ & $-1.8$ & $-5.0$ & $-1.6$ & $-5.0$ &$-1.9$\\
\hline
Significance of $M_c$
& & & & 
 & $2.6\sigma$ & $1.5\sigma$ & $2.7\sigma$ & $1.4\sigma$ & $2.7\sigma$ & $1.5\sigma$ \\
\hline
\end{tabular}
\label{table:ICs}
\end{table*}%
}

\subsection{Validation of the Systematics}
We now discuss and validate the robustness of our results to the presence of possible systematic errors and varying analysis choices. 

\begin{figure}[ht]
    \centering
    \includegraphics[width=0.48\textwidth]{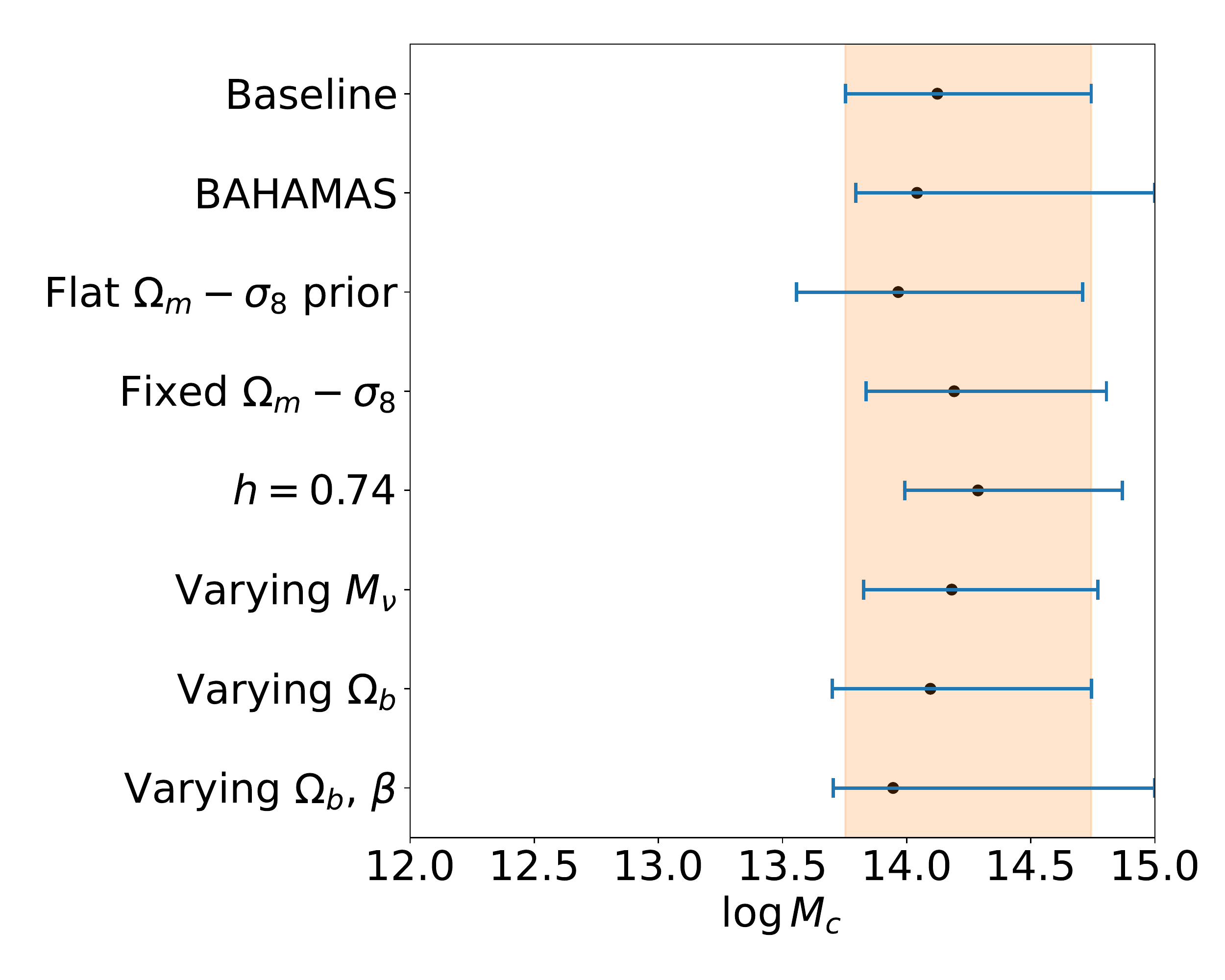}
    \caption{Systematic tests using real data. The x-axis spans the $\log{M_c}$ prior range. The top horizontal blue bar, which agrees with the vertical shaded region, is the marginalized $68\%$ C.L. constraint on $\log{M_c}$ in our baseline BCM analysis. Each subsequent horizontal error bar is the constraint on $M_c$ from an analysis with one alternative analysis choice relative to the baseline, as indicated in the legend. The `BAHAMAS' analysis fixes the baryonic parameters (other than $\log{M_c}$) to the best-fit values in the BAHAMAS\_nu0.06\_Planck2015 matter power spectrum at $z=0$ \cite{McCarthy2017}. The `Flat $\Omega_m$--$\sigma_8$ prior' analysis turns off the 2D Gaussian prior in the baseline analysis, varying these two parameters in the \texttt{Baccoemu} range with flat priors. The `Fixed $\Omega_m$--$\sigma_8$' chain fixes the values of these two parameters to their DES Year-3 3x2pt means, $\Omega_m = 0.339$ and $\sigma_8 = 0.733$. The `$h=0.74$' analysis fixes the Hubble parameter to the higher SH0ES value \cite{Riess:2019cxk} instead of the Planck value adopted by our baseline in Table \ref{table:pars}. The last three analyses, labeled as `Varying (parameter name)', apply flat priors to the corresponding parameters, in the range of \texttt{Baccoemu}. The parameter ranges for \texttt{Baccoemu} can be found at https://baccoemu.readthedocs.io/en/latest/.}
    \label{fig:mcsystematics}
\end{figure}

Figure \ref{fig:mcsystematics} shows the marginalized constraints on $M_c$. The top horizontal error bar corresponds to the baseline BCM analysis, while each subsequent error bar corresponds to an analysis with one alternative analysis choice relative to the baseline, as indicated in the legend. All of the alternative results agree with the baseline results to well within statistical errors. Interpreted in the context of our discussion  on the possible systematics in Section \ref{sec:systematics}, we conclude:
\begin{itemize}
    \item The agreement between the baseline result and the BAHAMAS, as well as the `Varying $\Omega_b, \beta$' result justifies our assumption to fix the baryonic feedback mode (constructed by OWLS-AGN simulation at $z=0$). In particular, our current measurement precision is not sensitive enough to distinguish this from the alternative BAHAMAS $z=0$ baryonic feedback mode, or else from the variation of the halo mass - gas fraction slope $\beta$. This result justifies our Fisher forecast in Section \ref{sec:fisher}. 
    
    The slight widening of the $M_c$ error bar is correlated with the negative $\beta$ BCM parameter in the BAHAMAS bestfit values, and the negative region allowed by the analysis varying $\beta$. The reason is that, given the gas fraction in the halo scaling as Equation \ref{eq:gasfraction}, positive $\beta$ suggests that the baryonic feedback is stronger toward less-massive halos, and vice versa. Recall that lower $f_{\rm gas}$ is a signature of stronger gas ejecting processes like AGN. Since the average halo mass of the DES galaxy sample is $\sim 10^{14}M_{\odot} $ \cite{DES:2018kma}, for fixed positive value of $\beta$, lower $M_c$ ($<10^{14}M_{\odot}$) suggest weaker baryonic feedback in the DES galaxy sample. However for a negative $\beta$ value, wider $M_c$ area in our prior range accommodates a substantial baryonic feedback for $\sim 10^{14}M_{\odot}$ population, so we get a wider error bar.
    
    The above reasoning further supports that the halo mass population in DES galaxy sample might have witnessed a substantial baryonic feedback. 
    
    \item The agreement between the baseline result and the `Flat $\Omega_m$--$\sigma_8$', `Fixed $\Omega_m$--$\sigma_8$', `$h=0.74$' and `Varying $M_{\nu}$' cases justifies our  assumptions to fix the cosmological parameters. In other words, these alternatives to our baseline cosmological model do not change our constraint on the baryonic parameter $M_c$. It is true that we cannot explore all of the possible changes to the fiducial cosmological parameters in these limited tests, as the \texttt{polychord} chains would have difficulty converging with too many unconstrained cosmological parameters. However, these single-parameter-change tests, along with the Fisher PCA forecast arguments in Section \ref{sec:fisher}, give us sufficient confidence that our detection of the baryonic feedback is not be due to bias in the standard cosmological parameters.
    
    \item We investigated our baseline posterior on the intrinsic alignment TATT model parameters, and the latter's degeneracy with $M_c$. The relevant constraints are shown in the Figure \ref{fig:tatt} in Appendix \ref{app:IA}, along with the constraint on the same set of TATT parameters from the DES Year-3 cosmic shear (1x2pt) and cosmic shear combined with galaxy clustering (3x2pt) analysis. The 2D contours in TATT parameters cross $M_c$ panels look highly uncorrelated between each other, suggesting that the scale-dependence of the IA signal (modeled by TATT) and baryonic suppression signal is fairly distinct. Thus the possibility that the potential degeneracy with the extended intrinsic alignment degrees of freedom causing the nontrivial $M_c$ constraint that deviates from its prior lower bound is also unlikely. 
    
\end{itemize}

The list of systematic checks just discussed is not guaranteed to be complete. In that regard, there are several caveats in our analysis that one should keep in mind:
\begin{itemize}
    \item Some systematics, for example the magnification, were argued to be small and were conventionally ignored in the previous work. However, the arguments and tests for such systematics were done at large scales that are relevant to the cosmological analysis \cite{krause2021dark}. It remains to be rigorously investigated whether these assumptions still apply at smaller scales that we use here. In contrast, other systematics, such as the Limber approximation and redshift-space-distortion effects,
    decrease when going to smaller scales, so we should be safe from them here. 
    \item The emulator sampling is limited in the model parameter space and wavenumber space. For example, our posterior on $\sigma_8$ is cut off at $0.73$  because \texttt{Baccoemu} only samples down to this value. Additionally, the nonlinear matter power spectrum sampling of \texttt{Baccoemu} goes up to $k = 5.0 h \rm Mpc^{-1}$, and beyond that wavenumber we need to extrapolate in order to compute the theory prediction for $\xi_\pm$. This limitation prevents us from modeling any enhancement of the matter power at smaller scales. We did however check, on several runs of the theory model, that including a high-k enhancement in power of roughly the expected  typical magnitude only introduces a small correction to the overall baryonic-effect $\Delta \chi^2$. For example, when we change the maximum wavenumber to which the baryonic suppression is applied from $k=5.0  \hmpcinv$ (which is the default in our analysis and incorporates no high-k enhancement) to $k=30.0 \hmpcinv$ (which is realized by the direct measurements from OWLS-AGN and DMO simulations so includes the enhancement effect), the two scenarios differ by only $\sim 5\%$ of the baseline $\Delta \chi^2$ difference between DMO and baryonic universe. 
    \item Baryonic feedback is a stochastic process, and in reality the baryon-corrected mass profile of halos may vary based on a number of physical properties of the halo ---  the halo age, formation history, etc. The baryon-correction model might not be able to capture all these dependencies. It is possible that the simplicity of our adopted baryonic correction model biases the baryonic parameter constraints. At the same time, it is unlikely that this simplicity induces a false detection of the baryonic suppression on the matter power spectrum because the baryonic effects become negligible for the current data precision, when $M_c \rightarrow 0$. 
    \item We assume $M_c$ to be constant with redshift. We note that X-ray observations of gas fractions in galaxy clusters are currently not accurate enough to provide a clear redshift trend \cite[see e.g.][and references therein]{hsc-xxl2021}, while hydrodynamical simulations predict different redshift dependences when varying subgrid physics \cite{Arico2020}. 
\end{itemize}

\begin{figure}[ht]
    \centering
    \includegraphics[width=0.48\textwidth]{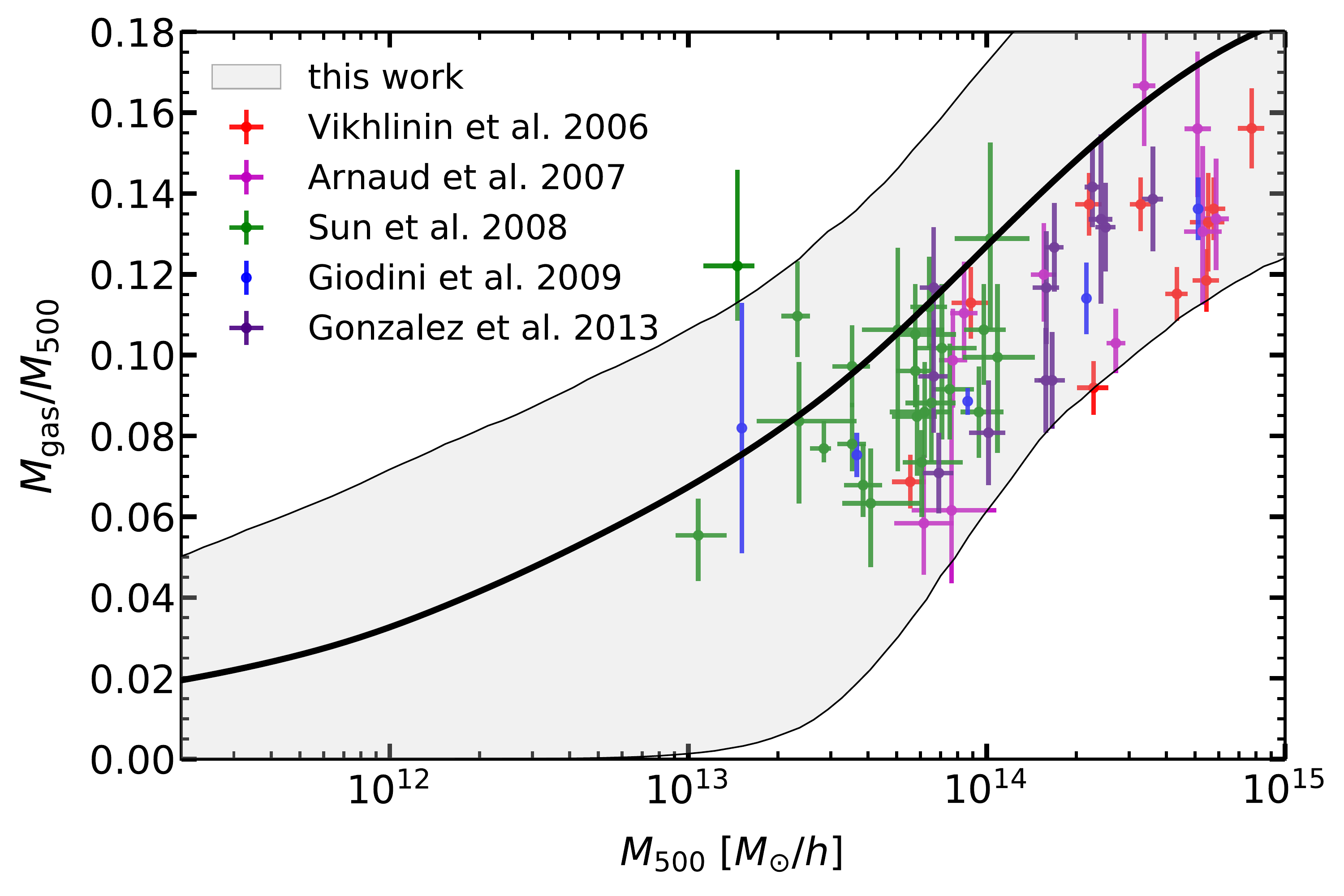}
    \caption{Gas mass fraction as a function of halo mass. The halo mass is computed assuming hydrostatic equilibrium. Data points are observations from different data sets, as reported in the legend. The grey shaded area highlights the 68\% credible region given by the constraints of baryonic parameters obtained in this work.}
    \label{fig:des_xray}
\end{figure}

\begin{figure}[ht]
    \centering
    \includegraphics[width=0.48\textwidth]{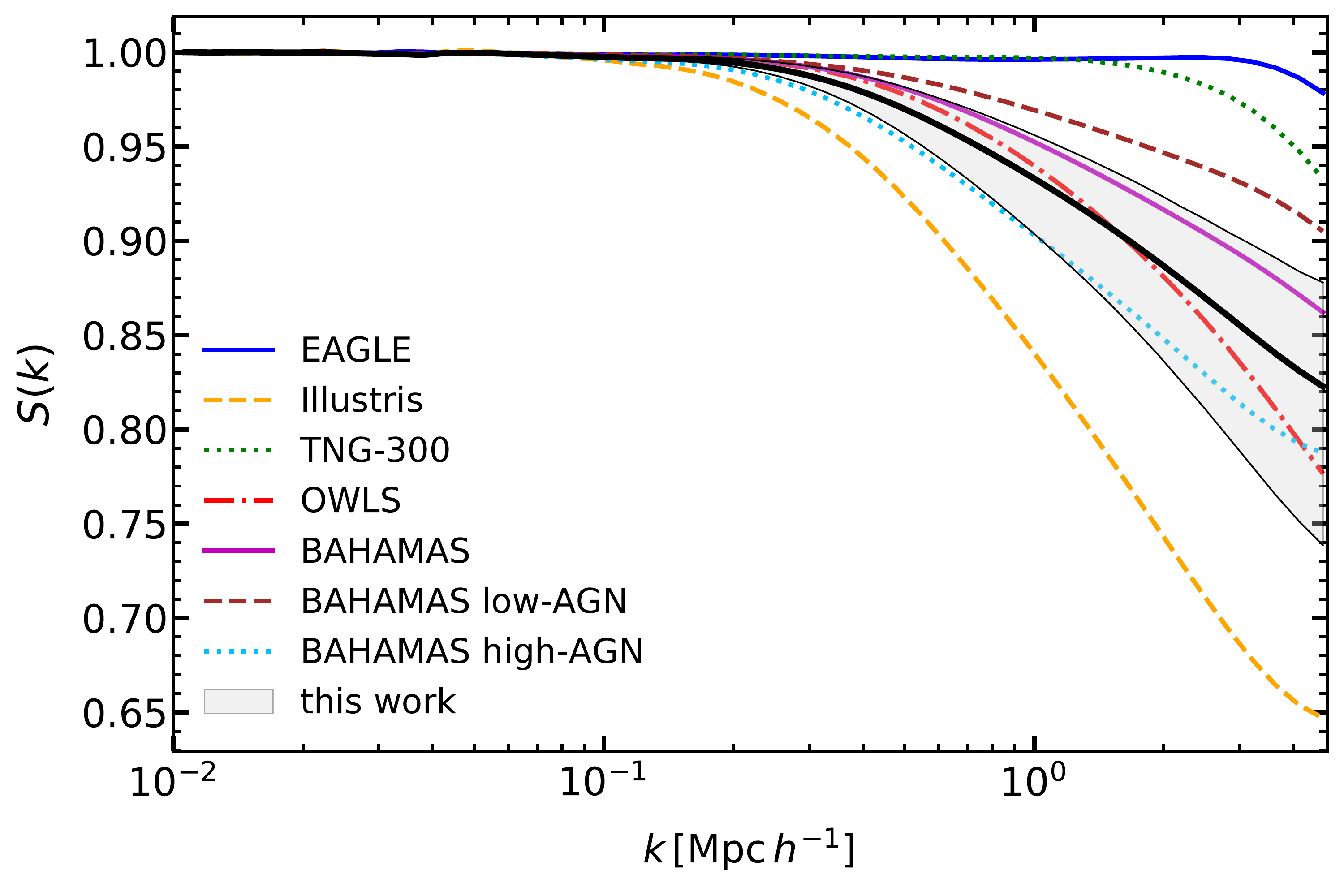}
    \caption{Suppression of the power spectrum due to baryonic effects, defined as the ratio $S(k,z=0)=P_{\rm baryons}/P_{\rm DMO}$.
    The lines show the suppression obtained when fitting different hydrodynamical simulations with the baryonic emulator at redshift $z=0$. The grey area highlights the 68\% credible region given by the constraints on the baryonic parameters obtained in this work. OWLS\_AGN seems to agree the most to our bestfit $S(k)$, however it should be borne in mind that we fix the baryonic parameters other than $M_c$ to OWLS\_AGN $z=0.0$ values in our analysis.}
    \label{fig:pk_hydro}
\end{figure}

\subsection{Comparison with X-ray Data and Previous Work}

The constraints on the baryonic parameters $M_c$ and $\beta$ that we have obtained can be directly translated to a prediction of the quantity of gas retained in haloes, through the Baryonic Correction Model. In Figure \ref{fig:des_xray} we compare this prediction to observations of the gas fractions in X-ray from \cite{Vikhlinin2006,Arnaud2007,Sun2009,Giodini2009,Gonzalez2013}. The mass of the haloes in these observations is obtained by assuming hydrostatic equilibrium, i.e. neglecting non-thermal contributions to the pressure. To fairly compare with our theoretical modeling, we rescale our halo masses by a factor $(1+b_{\rm h})$, where $b_{\rm h}$ is the so-called hydrostatic mass bias. We assume a Gaussian distribution of $b_{\rm h}$ with mean $0.26$ and standard deviation $0.07$, based on Ref.~\cite{HurierAngulo2018}. We show that the gas fractions directly observed are in good agreement with the 68\% credible region obtained from the cosmic shear. In particular, it appears that X-ray observations already have the potential to put tight constraints on baryonic parameters, opening up to joint constraints from lensing and X-ray, as done e.g. in Ref.~\cite{Schneider2021}. However, some complications may arise when joining different X-ray data sets, for instance when assessing their covariance or when marginalizing over the hydrostatic mass bias. Works such as Ref.~\cite{hsc-xxl2021}, which aims at building large homogeneous samples of clusters gas fractions over a wide range of halo masses, will be of great benefit in providing tighter constraints on baryonic parameters --- and thus in constraining the impact of the baryons on the matter power spectrum.

In Figure \ref{fig:pk_hydro} we show the baryonic suppression in the power spectrum that we expect at redshift $z=0$, given the constraints on the baryonic parameters, $M_c$ and $\beta$,  and universal baryon fraction, $\Omega_b/\Omega_m$,  obtained in this work. We compare the 68\% credible region given by our constraints with the power spectrum suppression predicted by different hydrodynamical simulations: EAGLE, Illustris, Illustris TNG, OWLS-AGN, BAHAMAS \cite{Schaye2015,Vogelsberger2014,Springel2018,Schaye2010,McCarthy2017}. The suppression that we find in this paper is compatible to that of the BAHAMAS simulations, particularly between in their versions with the medium- and high-temperature AGN feedback, and with OWLS-AGN. Note that BAHAMAS has been calibrated with the gas mass inside galaxy groups. We have thus shown that this BAHAMAS prediction is in a very good agreement with the gas fraction implied by our small-scale, cosmic shear analysis.         

Our analysis, which uses the small scales of DES Year-3 cosmic shear measurements, suggests a baryonic suppression $S(k)$ of the matter power spectrum $\approx 5\%$ at $k=1.0 \hmpcinv$ and $\approx 15 \%$ at $k=5.0 \hmpcinv$. Other previous work used weak lensing to constrain the baryonic feedback on matter power spectrum. Specifically, Ref.~\cite{DES:2020rmk} used DES Year-1 3x2pt measurements to constrain baryonic feedback  using principal components of the baryonic effect signature on the power spectrum as determined by numerical simulations. Because DES Year-1 measurements are less precise than Year-3, no conclusive constraint on the baryonic feedback was drawn at the time. More recent work in Ref.~\cite{Schneider2021} used KiDS-1000 \cite{KiDS:2020suj} as their weak lensing data set to constrain the baryonic feedback. While they could impose no informative constraint on their (seven-parameter) baryonic model, their derived effect on the matter power spectrum is broadly consistent with our results.
Ref.~\cite{Yoon:2020bop} compared the KiDS-450  measurements and the theory prediction by HMcode \cite{Mead:2016zqy} to find a substantially stronger baryonic feedback than what we and many AGN simulations find. However they have fairly large uncertainties, and only exclude the dark-matter-only case at $\sim 1.2 \sigma$. Recently, there has also been an effort in the community to measure baryonic feedback by combining weak lensing with thermal Sunyaev-Zeldovich signatures measured in CMB observations. Such an attempt with KiDS-1000 \cite{Troster:2021gsz} obtained baryonic constraints consistent with BAHAMAS simulation, and consequently in agreement with our findings as well.
Similarly, \cite{gatti2021cross,Pandey2021} have cross-correlated the cosmic shear measured by DES Year-3 with the Sunyaev-Zeldovich effect measured by Planck and ACT \cite{Planck_SZ_2016,ACT2020}, and modelling the signal with an hybrid approach based on hydrodynamical simulation and HMcode, finding hints of strong feedback compatible with Cosmo-OWLS high AGN \cite{LeBrun2014}, which is in broad agreement with BAHAMAS high-AGN.  
Ref.~\cite{chen2022thermal} used thermal Sunyaev-Zeldovich map from Planck around stacked DESI catalog to explore the baryonic feedback. They found $\beta$ to be always positive, which is consistent with our analysis choice of fixing $\beta=0.321$,; their constraint on $M_c$ also agrees with our findings. 

In summary, a number of earlier analyses that constrained baryonic feedback found results that are  consistent with our ours.

In closing this Section, we note that our results are based on a straightforward analysis that uses solely the DES Y3 cosmic shear measurements, and has been subjected to a battery of systematic tests.  Because of the conservative assumptions that we made,  the preference we find for the baryonic suppression, while not statistically overwhelming (at $1.4 \sigma - 2.7 \sigma$, depending on the assumptions), is robust.


\section{Conclusions}\label{sec:concl}

In this paper we constrain the effect of baryonic feedback on the matter power spectrum. As a starting point, we adopt the baryon correction model (BCM) \citep{SchneiderTeyssier2015, Arico2020} which introduces seven parameters to model the baryon corrected halo mass profile. We choose to fix all cosmological parameters except $\Omega_m$ and $\sigma_8$, and focus our attention on the baryonic sector. Specifically, we use only small angular scales in DES cosmic shear measurements to constrain the baryonic feedback. Our analysis is therefore complementary to the standard cosmological analysis that discards the  small scales that we are using here, and instead uses large scales to constrain cosmology (and largely avoid the effect of baryons).

We demonstrate by means of a Fisher forecast that our DES Year-3 small-scale cosmic shear measurements are sensitive enough to constrain only one BCM parameter, $\log{M_c}$, where $M_c$ is a typical mass scale related to the gas content of halos. We also carry out a battery of tests to validate our results, specifically studying the impact of alternative assumptions in the choice of priors, parameters that are fixed or varied, and alternative models for nonlinear dark-matter clustering.

We  constrain the baryonic parameter $\log{M_c}$ to be $14.12^{+0.62}_{-0.37}$ at $68\% \rm C.L.$, while fixing other baryonic (BCM) parameters to the bestfit of OWLS-AGN hydrodynamic simulation. Our analysis prefers the best-fit baryonic model to the best-fit dark-matter-only alternative (which corresponds to $\log{M_c}=12.0$ in our analysis) at the $\sim\!\! 2 \sigma$ significance. 

We find good agreement between our cosmic-shear constraints on the baryonic feedback and independent X-ray measurements, as illustrated in Figure \ref{fig:des_xray}. This result foreshadows exciting future possibilities: one could use independent X-ray, thermal Sunyaev-Zeldovich effect, and other observations as a prior on the baryonic-feedback parameter space, in turn enabling more precise constraints on the latter. We hope to incorporate this approach in the future, and combine it with the forthcoming DES Year-6 cosmic-shear data.  

\section*{Acknowledgements}
Giovanni Aric\`{o} and Dragan Huterer thank the Max Planck Institute for Astrophysics for hospitality, where conversations that led to this project originated.

Funding for the DES Projects has been provided by the U.S. Department of Energy, the U.S. National Science Foundation, the Ministry of Science and Education of Spain, 
the Science and Technology Facilities Council of the United Kingdom, the Higher Education Funding Council for England, the National Center for Supercomputing 
Applications at the University of Illinois at Urbana-Champaign, the Kavli Institute of Cosmological Physics at the University of Chicago, 
the Center for Cosmology and Astro-Particle Physics at the Ohio State University,
the Mitchell Institute for Fundamental Physics and Astronomy at Texas A\&M University, Financiadora de Estudos e Projetos, 
Funda{\c c}{\~a}o Carlos Chagas Filho de Amparo {\`a} Pesquisa do Estado do Rio de Janeiro, Conselho Nacional de Desenvolvimento Cient{\'i}fico e Tecnol{\'o}gico and 
the Minist{\'e}rio da Ci{\^e}ncia, Tecnologia e Inova{\c c}{\~a}o, the Deutsche Forschungsgemeinschaft and the Collaborating Institutions in the Dark Energy Survey. 

The Collaborating Institutions are Argonne National Laboratory, the University of California at Santa Cruz, the University of Cambridge, Centro de Investigaciones Energ{\'e}ticas, 
Medioambientales y Tecnol{\'o}gicas-Madrid, the University of Chicago, University College London, the DES-Brazil Consortium, the University of Edinburgh, 
the Eidgen{\"o}ssische Technische Hochschule (ETH) Z{\"u}rich, 
Fermi National Accelerator Laboratory, the University of Illinois at Urbana-Champaign, the Institut de Ci{\`e}ncies de l'Espai (IEEC/CSIC), 
the Institut de F{\'i}sica d'Altes Energies, Lawrence Berkeley National Laboratory, the Ludwig-Maximilians Universit{\"a}t M{\"u}nchen and the associated Excellence Cluster Universe, 
the University of Michigan, NSF's NOIRLab, the University of Nottingham, The Ohio State University, the University of Pennsylvania, the University of Portsmouth, 
SLAC National Accelerator Laboratory, Stanford University, the University of Sussex, Texas A\&M University, and the OzDES Membership Consortium.

Based in part on observations at Cerro Tololo Inter-American Observatory at NSF's NOIRLab (NOIRLab Prop. ID 2012B-0001; PI: J. Frieman), which is managed by the Association of Universities for Research in Astronomy (AURA) under a cooperative agreement with the National Science Foundation.

The DES data management system is supported by the National Science Foundation under Grant Numbers AST-1138766 and AST-1536171.
The DES participants from Spanish institutions are partially supported by MICINN under grants ESP2017-89838, PGC2018-094773, PGC2018-102021, SEV-2016-0588, SEV-2016-0597, and MDM-2015-0509, some of which include ERDF funds from the European Union. IFAE is partially funded by the CERCA program of the Generalitat de Catalunya.
Research leading to these results has received funding from the European Research
Council under the European Union's Seventh Framework Program (FP7/2007-2013) including ERC grant agreements 240672, 291329, and 306478.
We  acknowledge support from the Brazilian Instituto Nacional de Ci\^encia
e Tecnologia (INCT) do e-Universo (CNPq grant 465376/2014-2).

This manuscript has been authored by Fermi Research Alliance, LLC under Contract No. DE-AC02-07CH11359 with the U.S. Department of Energy, Office of Science, Office of High Energy Physics.

\bibliographystyle{ieeetr}
\bibliography{baryonification}

\appendix 
\section{Baryon Correction Model}\label{app:BCM}
The Baryon Correction Model (BCM), also known as \textit{baryonification} \cite{SchneiderTeyssier2015,Arico2020}, is a scheme to perturb the output of N-body simulations to include given baryon processes. Each halo in the simulation is decomposed into a dark matter and baryonic component with a respective density profile associated. The difference between the profiles is then used to compute a displacement field that is applied to the particles of the halo. The functional forms of the density profiles are motivated by observations, theoretical arguments, and hydrodynamical simulations, and they depends on a few free parameters.\\
The scheme we use in this work decomposes the halo in dark matter, gas, and galaxies \cite{Arico2020b}. The gas can be bound to its halo in hydrostatic equilibrium, ejected by some feedback process, or reaccreted, whereas the galaxies can be central or satellites. The baryonic gravitational potential back-reacts onto the dark matter, causing a quasi-adiabatically relaxation. 
The evaluation of the baryonic effects on the power spectrum are speed up by using a neural network emulator \cite{Arico2020c}. 
This model has a total of 7 free parameters, but in Section \ref{sec:fisher} we show that our data is mostly sensitive to one parameter, namely $M_c$. This parameter regulates the amount of gas that is retained in halos, $f_{\rm gas}$, and therefore also the quantity of gas ejected by baryonic feedback, through the equation
\begin{equation}
    f_{\rm gas} = \frac{\Omega_b/\Omega_m - f_{\rm gal}}{1+(M_c/M_{200})^{\beta}}, 
    \label{eq:gasfraction}
\end{equation}
where $f_{\rm gal}$ is the mass fraction of galaxies, $M_{200}$ is the total mass of the halo, and $\beta$ another free parameter. Therefore, $M_c$ is defined as the characteristic halo mass for which half of the halo gas is depleted.\\
The parameters $\theta_{\rm inn}$, $\theta_{\rm out}$, and $M_{\rm inn}$ regulate the density of the hydrostatic gas in the halo, according to a double power-law shape. The parameter $\eta$ set the maximum distance from the halo that the ejected gas can reach, in units of the halo escape radius. Finally, $M_{\rm z0,cen}$ regulates the characteristic galaxy mass fraction following an abundance matching scheme. We refer to \cite{Arico2020,Arico2020b} for the complete equations and baryonic functional forms of the BCM employed.\\
In this work, we fix all the parameters except $M_c$ to the best-fit values obtained fitting the OWLS-AGN hydrodynamical simulation in \cite{Arico2020c}.

\section{Intrinsic Alignment Parameters}\label{app:IA}

The measurement of the averaged cosmic shears is based on the directly obtained ellipticities of the galaxies. However, because the galaxies are formed in the gravitational field of the large scale structures, they have non-spherically randomly distributed shapes (intrinsic alignments) under the effect of tidal forces. The intrinsic alignment is another major source of astrophysical systemtatics at small scales of the cosmic shear measurement, other than the baryonic suppression we are studying in this work. Hence we are obliged to investigate whether any IA signals would be degenerate with the baryonic suppression, thus resulting in a fake detection of the baryonic suppression. Although a reliable modeling of the IA terms at small scale is not currently available to the best of our knowledge, Figure \ref{app:IA} indicate two facts: 1. There is no strong correlation between baryonic parameter $M_c$ and IA-TATT parameters $A_1A_2, A_1/A_2, \alpha_1, \alpha_2$ and $\rm bias_{\rm ta}$; 2. The IA-TATT model \cite{Blazek:2017wbz} parameters constraints for our baseline analysis using only the small scales of the cosmic shear, are consistent with DES Year-3 large scale cosmic shear 1x2pt and cosmic shear + clustering 3x2pt analysis.

Given the $M_c$-IA parameters contour plots, we conclude that the IA signal is not substantially correlated with the baryonic suppression pattern, so will not introduce significant systematics to our baryonic constraints. What is more, by comparing with other constraints on the IA parameters from DES Year-3 1x2pt and 3x2pt analysis, the consistent IA results show that there is indeed no unexpected IA signal in our small scales cosmic shear analysis.

\section{Effective Redshift of the BCM Constraint}\label{app:effectiveredshift}
As a photometric survey, DES galaxy catalogs do not have high precision measurements on the redshifts, so the astrophysical and cosmological findings usually contain information blending in a range of redshifts. However, in the Section \ref{sec: results}, we provide the effective redshift at which our constraint on $M_c$ is attached to. In this section we explain how we get this number $z_{\rm eff}=0.21$. 

The strategy is based on the following reasoning: the redshift-localized baryonic suppression effect that makes the most difference in the statistics (likelihood, or $\chi^2,$) is the redshift our measurement most sensitive to, in terms of the baryonic suppression effect constraints. So we apply a Gaussian kernel with width $\sigma_z = 0.1$ on the baryonic suppression $S(k,z)$ modeled by \texttt{Baccoemu}. At the center redshift, BCM $S(k,z)$ is multiplied to the dark matter only matter power spectrum, while away from the Gaussian kernel center $S(k,z) \rightarrow 1$. Scanning $z_{\rm center}$ through from 0.0 to 1.0, we find the $\Delta \chi^2$ between redshift-localized BCM and DMO data vector peaks at 0.21. Thus we conclude $z_{\rm eff}=0.21$ is the effective redshift contributing the most to our baryonic feedback constraining power.

\begin{figure}[ht]
    \centering
    \includegraphics[width=0.48\textwidth]{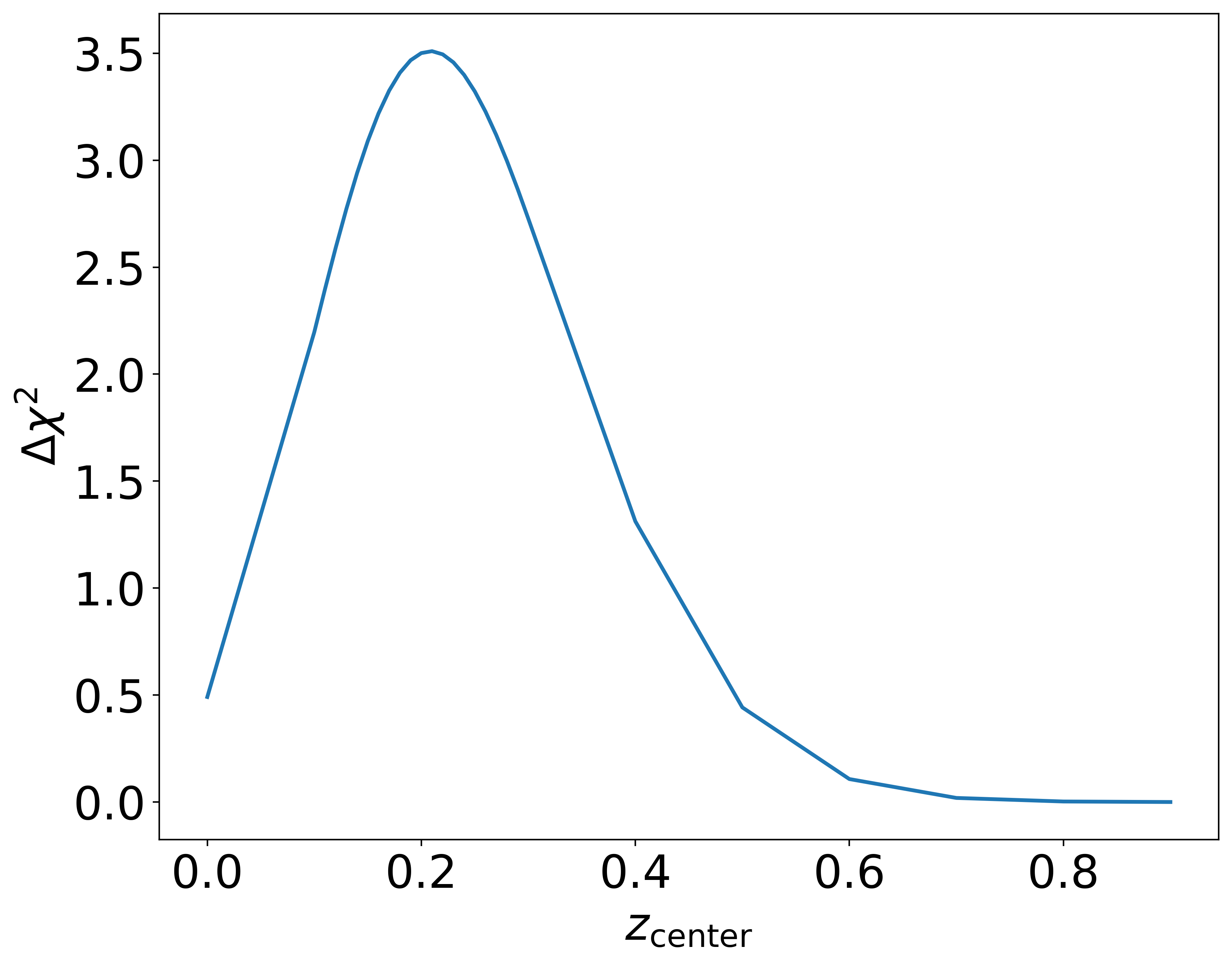}
    \caption{The $\Delta \chi^2$ between redshift-localized BCM and DMO data vectors. The localization of BCM baryonic suppression $S(k,z)$ is implemented by Gaussian kernels with $\sigma_{z}=0.1$.}
    \label{fig:zeff}
\end{figure}

\section{Information Criteria Definitions}\label{app:ICs}

It is a inclusive debate that which information criterion serves the best (un-biased and statistically significant) for the purpose of model comparison in cosmology. However, if a finding is significant enough, we believe it should show up regardless of the metric, so we present all the popular metrics in the result section for readers to choose their favourite. Table \ref{table:ICdefinitions} gives the unambiguous definitions of the information criteria we presented in the main text.

{\renewcommand{\arraystretch}{1.3}
\begin{table}[ht]
\centering
\caption{Information Criteria Definitions. Our data $\bold{D}$ consists of the small scale cosmic shear $\xi_{\pm}$ measurements from DES Year-3, with the number of data points $N_{\rm pts} = 173$. All averages are done by integrating the posterior, namely the average by weight of the Monte Carlo chain.}
\begin{tabular}[t]{cc}
\hline
Name of the Quantity & Formula\\
\hline
$\chi^2$ & $(\bold{M}-\bold{D})^T C^{-1}(\bold{M}-\bold{D})$\\
\hline
BMD & \makecell{$\langle (-\frac{\chi^2}{2}-\mathcal{Z})^2 \rangle - \langle -\frac{\chi^2}{2}-\mathcal{Z}\rangle^2$ \\ $\mathcal{Z}$ being the logarithm evidence} \\
\hline
AIC & $\chi^2_{\rm min}+2*\rm BMD$\\
\hline
AIC($k=N$) & $\chi^2_{\rm min} + 2*N_{\rm model}$ \\
\hline 
BIC & $\chi^2_{\rm min} + {\rm BMD}*\log(N_{\rm pts})$\\
\hline
BIC($k=N$) & $\chi^2_{\rm min} + N_{\rm model}*\log(N_{\rm pts})$\\
\hline
DIC & $\langle\chi^2\rangle + 2*{\rm BMD}$\\
\hline
DIC($k=N$) & $\langle\chi^2\rangle + 2*N_{\rm model}$\\
\hline
\end{tabular}
\label{table:ICdefinitions}
\end{table}%
}

\begin{figure*}[ht]
    \centering
    \includegraphics[width=0.95\textwidth]{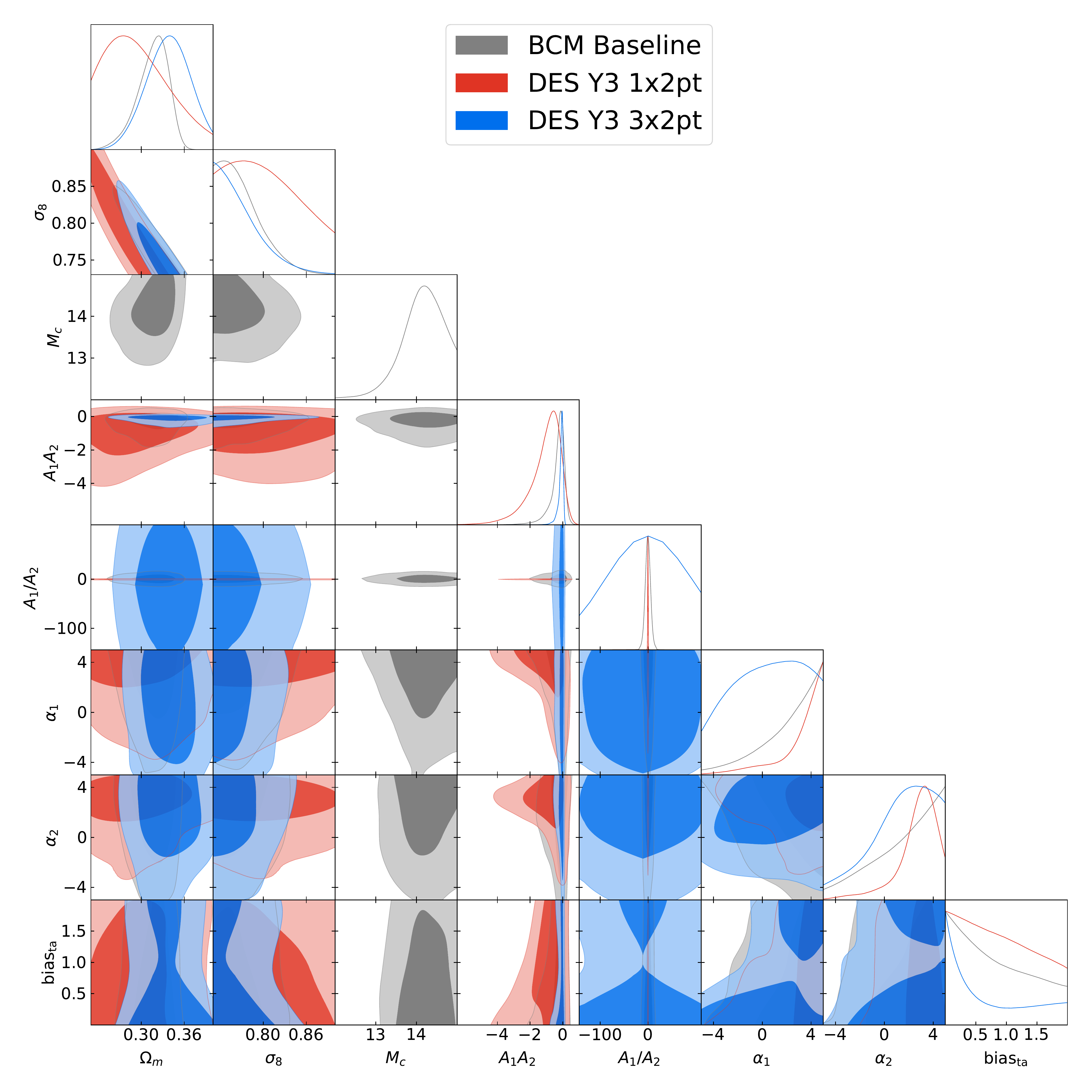}
    \caption{Contour plots for $\Omega_m, \sigma_8$ and $M_c$, from BCM baseline analysis MCMC chain and DES Year-3 1x2pt and 3x2pt \cite{DES:2021wwk} MCMC chains.}
    \label{fig:tatt}
\end{figure*}

\clearpage
\end{document}